\documentclass[aps,prl,reprint,superscriptaddress]{revtex4-2}

\usepackage{graphicx}
\usepackage{calligra}
\usepackage{upgreek}
\usepackage{color}
\usepackage{inputenc,caption}
\usepackage{stackrel}
\usepackage{subcaption}
\usepackage{placeins}
\usepackage{amsmath}
\usepackage{soul}
\usepackage{amssymb}

\DeclareMathAlphabet{\mathpzc}{OT1}{pzc}{m}{it}

\newcommand{\be}{\begin{equation}}
\newcommand{\ee}{\end{equation}}
\newcommand{\bea}{\begin{eqnarray}}
\newcommand{\eea}{\end{eqnarray}}
\newcommand{\lb}{\label}

\newcommand{\bv}{{\bf v}}
\newcommand{\bu}{{\bf u}}

\newcommand{\bk}{{\bf k}}

\newcommand{\bx}{{\bf x}}

\newcommand{\br}{{\bf r}}

\newcommand{\boeta}{{\mbox{\boldmath $\xi$}}}

\newcommand{\grad}{{\mbox{\boldmath $\nabla$}}}
\newcommand{\bdot}{{\mbox{\boldmath $\cdot$}}}

\newcommand{\uit}{\mathpzc{u}}

\newcommand{\bLambda}{{\mbox{\boldmath $\Lambda$}}}

\newcommand{\nocontentsline}[3]{}
\newcommand{\tocless}[2]{\bgroup\let\addcontentsline=\nocontentsline#1{#2}\egroup}

\begin{document}

\definecolor{orange}{RGB}{255,165,0}

\setlength{\abovedisplayskip}{8pt}
\setlength{\belowdisplayskip}{8pt}

\title{Spontaneous stochasticity amplifies \textcolor{black}{even} thermal noise to the largest scales of turbulence in a few eddy turnover times}

\author{Dmytro Bandak}
\affiliation{Department of Physics, University of Illinois at Urbana-Champaign, Loomis Laboratory of Physics, 1110 West Green Street, Urbana, Illinois 61801, USA}
\author{Alexei Mailybaev}
\affiliation{Instituto de Matem\'atica Pura e Aplicada - IMPA, Rio de Janeiro, Brazil}
\author{Gregory L. Eyink}
\affiliation{Department of Applied Mathematics \& Statistics, and Department of Physics \& Astronomy, The Johns Hopkins University, Baltimore, MD, USA, 21218}
\author{Nigel Goldenfeld}
\affiliation{Department of Physics, University of California, San Diego, 9500 Gilman Drive, La Jolla, California 92093, USA}

\begin{abstract}
%Thermal noise is inevitably present in molecular fluids and significantly influences the dissipation range of turbulent flows.  Do any effects of thermal noise extend to larger scales?  Here we use theoretical estimates and shell model simulations to argue that Eulerian spontaneous stochasticity, a manifestation of the non-uniqueness of the solutions to the Euler equation that is conjectured to occur in Navier-Stokes turbulence at high Reynolds numbers, leads to universal statistics at finite times, not just at infinite time as for standard chaos. We show that thermal noise effects vanish slowly enough with increasing Reynolds number that they are able to trigger spontaneous stochasticity.  If confirmed for Navier-Stokes turbulence, our findings would imply that intrinsic stochasticity of turbulent fluid motions at all scales can be triggered by unavoidable molecular noise, with implications for modeling in engineering, climate, astrophysics and cosmology.

How predictable are turbulent flows?  
Here we use theoretical estimates and shell model simulations to argue that Eulerian spontaneous stochasticity, a manifestation of the non-uniqueness of the solutions to the Euler equation that is conjectured to occur in Navier-Stokes turbulence at high Reynolds numbers, leads to universal statistics at finite times, not just at infinite time as for standard chaos. These universal statistics are predictable, even though individual flow realizations are not.
\textcolor{black}{Any small-scale noise vanishing} slowly enough with increasing Reynolds number 
\textcolor{black}{can} trigger spontaneous stochasticity \textcolor{black}{and here we show that 
thermal noise alone, 
in the absence of any larger disturbances, would suffice.} 
If confirmed for Navier-Stokes turbulence, our findings would imply that intrinsic stochasticity of turbulent fluid motions at all scales can be triggered \textcolor{black}{even}
by unavoidable molecular noise, with implications for modeling in engineering, climate, astrophysics and cosmology.
\end{abstract}

\maketitle

%\textit{Introduction.} 

Spontaneous stochasticity is a recently discovered phenomenon \cite{bernard1998slow,E2000generalized,thalabard2020butterfly} in turbulent flows \cite{frisch1995turbulence,SREE99},
whereby solutions of model fluid equations remain unpredictable and stochastic due to diverging Lyapunov exponents in the high Reynolds number limit, even though random perturbations to the flow are negligible asymptotically.  
It is an open question as to whether or not it occurs in real fluids or in the Navier-Stokes equations. 
%One source of tiny randomness is thermal noise arising from molecular agitation; it is inevitably present, but, despite the pioneering arguments of Betchov \cite{betchov1957fine, betchov1961thermal,betchov1964measure} and Ruelle \cite{ruelle1979microscopic}, its relevance to turbulence has been largely dismissed. 
%Here we argue that at large (but finite) Reynolds numbers, a stochastic wave propagating from small to large scales, as first postulated by Lorenz \cite{lorenz1969predictability}, rapidly randomizes the large-scale flow, \textit{even if the only sources of noise to trigger the wave are molecular fluctuations at small scales}.  We show that flow fluctuations at large-scales exhibit universal statistics due to spontaneous stochasticity, not to thermal fluctuations, which are negligible at large scales.  
Here we report that at large (but finite) Reynolds numbers, a stochastic wave propagating from small to large scales, as first postulated by Lorenz \cite{lorenz1969predictability}, rapidly randomizes the large-scale flow, \textit{even if the only sources of noise to trigger the wave are molecular fluctuations at small scales}.  Going beyond Lorenz, we show that flow fluctuations \textcolor{black}{{\it at large-scales} exhibit universal statistics due to spontaneous stochasticity and not directly due to whatever {\it small-scale} noise triggers the stochastic wave.  Spontaneous stochasticity is not inevitable; for it to be triggered, the small-scale noise must become negligible in the large Reynolds number limit sufficiently slowly.
The surprise is that even thermal noise satisfies this condition.
}
%At small scales, however,  thermal fluctuations are strong enough to trigger spontaneous stochasticity. 
These new indirect effects at large scales are distinct from a growing body of work showing that thermal noise directly alters the turbulent dissipation range below the Kolmogorov scale
\cite{betchov1957fine, betchov1961thermal,betchov1964measure,ruelle1979microscopic,bandak2021thermal,bandak2022dissipation,mcmullen2022navier, bell_nonaka_garcia_eyink_2022,ma2023effect,mcmullen2023thermal}, \textcolor{black}{and in fact other small-scale disturbances, typically much larger than thermal agitation, will produce indistinguishable large-scale stochasticity.}
%
%Eulerian spontaneous stochasticity, defined precisely below, rapidly randomizes the flow at all scales from the bottom of the so-called inertial range (where scaling laws are well-established \cite{frisch1995turbulence,SREE99}) up to the scale of the largest eddies, \textcolor{red}{even if the only source of noise is molecular fluctuations at small scales}.  
%
To demonstrate our claim that the probability distributions of relevant flow quantities have a universal, non-delta-function form at large but finite Reynolds numbers, we are forced to employ a simplified but well-studied dynamical model \cite{l1998improved} of  turbulence, since the Navier-Stokes equations are computationally unfeasible at the large Reynolds numbers required for convincing convergence to universal statistics. 
%to make a convincing demonstration of our claim, 
These results suggest an essential indeterminism of turbulent flows at scales of practical interest, with potentially far-ranging implications for engineering, geophysics, and astrophysics.

%\textcolor{green}{To demonstrate our claim, we use numerics to show directly that the probability distributions of relevant flow quantities have a universal, non-delta-function form at large but finite Reynolds numbers. The Navier-Stokes equations are computationally unfeasible at the large Reynolds numbers required for convincing convergence to universal statistics,
%to make a convincing demonstration of our claim, 
%so we are forced to employ a simplified but well-studied dynamical model \cite{l1998improved} of  turbulence. These results suggest an essential indeterminism of turbulent flows at scales of practical interest, with potentially far-ranging implications for engineering, geophysics, and astrophysics.}

%Recent experimental advances may permit observation of such effects \cite{van2022dispersion}.

\textit{Fluctuating Hydrodynamics.} The fluctuating hydrodynamics of Landau-Lifschitz \cite{landau1959fluid} describes the effect of thermal noise in fluid flows by including fluctuating stresses into the Navier-Stokes equation. It is expressed in a form non-dimensionalized by large-scale velocity $U$ and outer/integral length $L$ as

\vspace{-20pt}
\be \partial_t\bu + (\bu\cdot\grad)\bu = -\grad p + \frac{1}{Re}\triangle \bu + \sqrt{\Theta} \grad\cdot \boeta + F \bold{f}, \lb{FNS2} \ee
where the fluctuating stress is modeled as a Gaussian random field $\boeta$ with mean zero and covariance

\vspace{-20pt}
\begin{eqnarray}
\langle \xi_{ij}(\bx,t) \xi_{kl}(\bx',t')\rangle& = &
\left(\delta_{ik}\delta_{jl}+\delta_{il}\delta_{jk} -\frac{2}{3}\delta_{ij}\delta_{kl}\right)\cr
&& \hspace{20pt} \times \delta^3(\bx-\bx')\delta(t-t') \lb{FDR}.
\end{eqnarray}
Here, Reynolds number is defined as $Re \equiv UL/\nu$ where $U$ is the large-scale velocity of the flow, $L$ is the forcing 
scale at which energy is injected, and $\nu$ is kinematic viscosity.
$\Theta\equiv 2 \nu k_B T/ \rho L^4 U^3$ appears due to the presence of noise and characterizes its strength according to the fluctuation-dissipation relation. $k_B$ is Boltzmann's constant, $T$ is absolute temperature, and $\rho$ is mass density. To drive turbulence, we have added an external forcing $\bold{f}$ non-dimensionalized by 
r.m.s. value $f_{rms},$ with prefactor $F = Lf_{rms}/U^2$ setting the magnitude. Strictly speaking, equation (\ref{FNS2}) describes a coarse-grained velocity field on lengthscale $\Lambda^{-1}$ larger than the mean free path and some care is needed to interpret this mathematically, see Supplemental Material (SM \S I),  for details \cite{forster1977large,zubarev1983statistical,morozov1984langevin,espanol2009microscopic}.

% Slightly shortening this paragraph and adding a comment about the difference between steady state statistics and predictability

%If we use as flow parameters the values characteristic of the atmospheric boundary layer (ABL) \cite{garratt1994atmospheric} $T=300^\circ \ K$, $\nu = 1.5 \times 10^{-5} \ m^2/sec$, $\rho=1.2 \ kg/m^3$, $\varepsilon=4 \times 10^{-2} \ m^2/sec^3$, $L=10^3 \ m$, $U= 3.42 \ m/sec$,  where $\varepsilon$ is the mean energy dissipation per mass, we arrive at an exceedingly small value for $\Theta \simeq 2.59 \times 10^{-39}$. The smallness of $\Theta$ naively justifies dropping the fluctuating stress term from the Landau-Lifschitz equations, recovering the conventional point of view that the physically relevant turbulent fluid flows are well modeled by the deterministic (cut-off) Navier-Stokes equations. This conclusion is consistent with numerical findings \cite{bandak2022dissipation,bell_nonaka_garcia_eyink_2022,mcmullen2022navier} that the turbulent steady-state statistics at scales larger than the Kolmogorov length $\eta=\nu^{3/4}\varepsilon^{-1/4}$ are unchanged by molecular fluctuations.

Using flow parameters characteristic of the atmospheric boundary layer (ABL) \cite{garratt1994atmospheric} $T=300^\circ \ K$, $\nu = 1.5 \times 10^{-5} \ m^2/sec$, $\rho=1.2 \ kg/m^3$, $\varepsilon=4 \times 10^{-2} \ m^2/sec^3$, $L=10^3 \ m$, $U= 3.42 \ m/sec$,  where $\varepsilon$ is the mean energy dissipation per mass, we find $\Theta \simeq 2.59 \times 10^{-39}$, naively justifying dropping the fluctuating stress term from the Landau-Lifschitz equations and supporting the conventional wisdom that physically relevant turbulent fluid flows are well modeled by the deterministic (cut-off) Navier-Stokes equations. This conclusion is consistent with numerical findings \cite{bandak2022dissipation,bell_nonaka_garcia_eyink_2022,mcmullen2022navier} that the turbulent steady-state statistics at scales larger than the Kolmogorov length $\eta=\nu^{3/4}\varepsilon^{-1/4}$ are unchanged by molecular fluctuations, but not addressing the issue of the flow predictability.

% Slightly shortening the paragraph below and adding comment about predictability

% Besides the small amplitude of thermal noise, fully-developed fluid turbulence is defined by high Reynolds number, which for the ABL parameters amounts to $Re \simeq 2.28 \times 10^8$. Again, this naively justifies dropping the viscous term proportional to $Re^{-1}$ from the equations \cite{drivas2019remarks}. However, the limiting problem with no viscosity, no wavenumber cutoff, and no thermal noise are the continuum Euler equations, which do not have unique solutions and are formally ill-posed \cite{de2010admissibility,de2020non,daneri2021non}. Such non-uniqueness or ``flexibility'' of solutions suggests an intrinsic unpredictability of turbulent fluid motions at high Reynolds numbers due to \textit{Eulerian spontaneous stochasticity} \cite{mailybaev2016spontaneously,thalabard2020butterfly}, and provides a possibility for tiny thermal noise to influence all scales of the flow up to the largest.

Fully-developed fluid turbulence in the ABL has $Re \simeq 2.28 \times 10^8$, justifying dropping as well the viscous term proportional to $Re^{-1}$ from the equations \cite{drivas2019remarks}. However, the limiting equations with no viscosity, no wavenumber cutoff, and no thermal noise are the continuum Euler equations, which do not have unique solutions and are formally ill-posed \cite{de2010admissibility,de2020non,daneri2021non}. Such non-uniqueness or ``flexibility'' of solutions suggests an intrinsic unpredictability of turbulent fluid motions at high Reynolds numbers known as \textit{Eulerian spontaneous stochasticity} \cite{mailybaev2016spontaneously,thalabard2020butterfly}, and provides a possibility for tiny thermal noise to influence the predictability of all scales of the flow up to the largest.

\begin{figure*}
\centering
\begin{subfigure}{.31\linewidth}
\includegraphics[width=\linewidth]{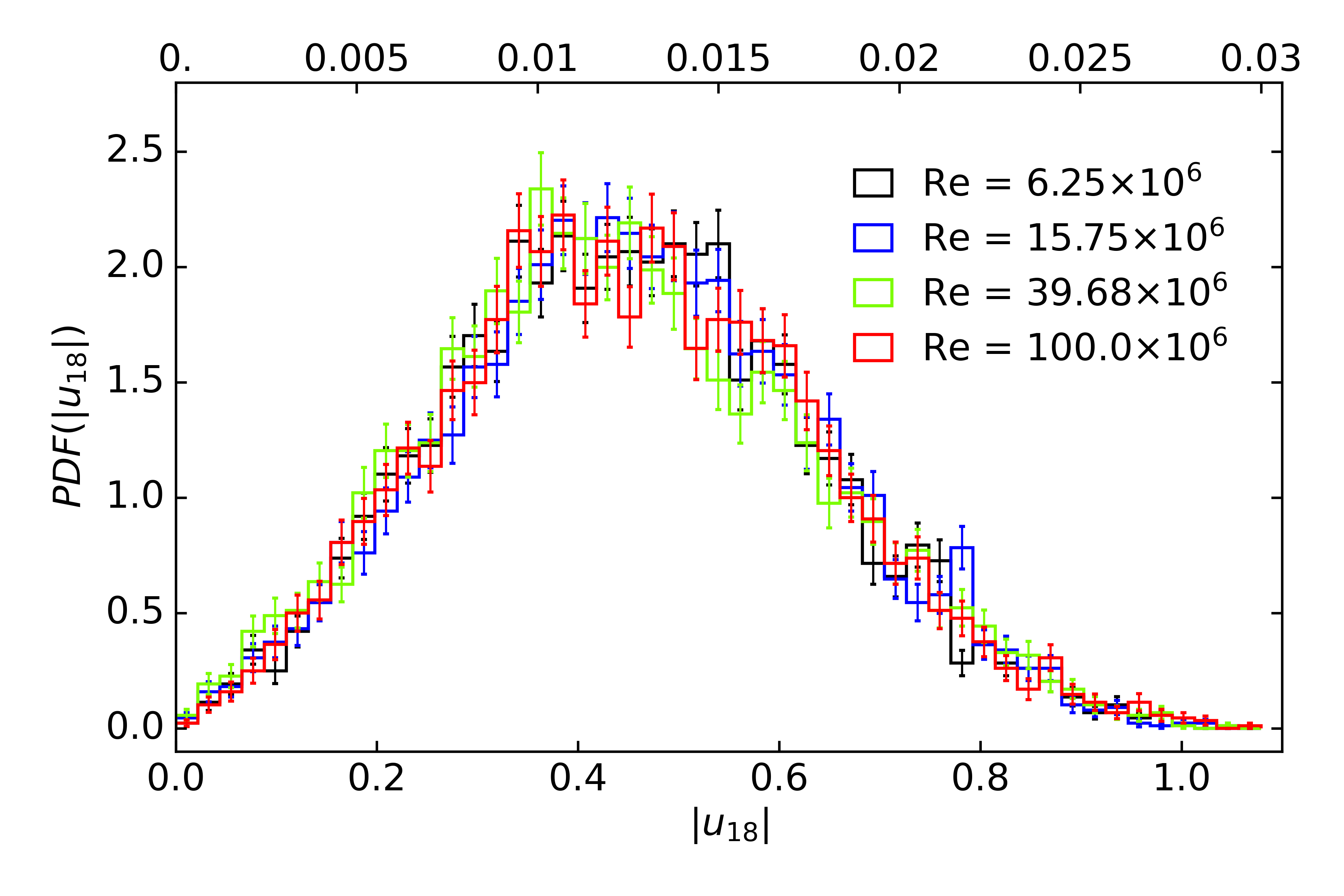}
\caption{Self-similar initial state with noise scaling $\alpha = 0.$}
\end{subfigure}
\begin{subfigure}{.31\linewidth}
\includegraphics[width=\linewidth]{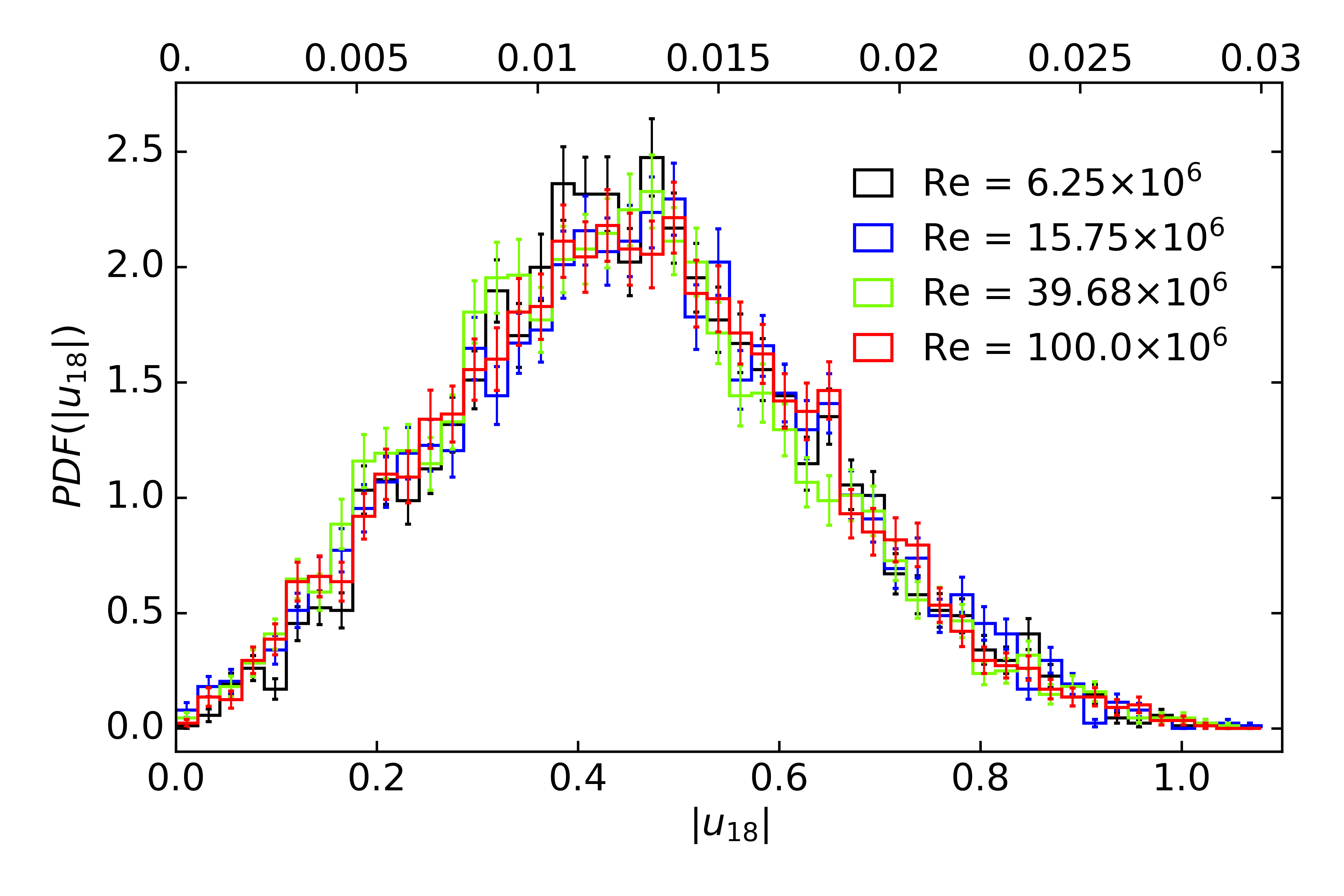}
\caption{Self-similar initial state with noise scaling $\alpha = 3.$}
\end{subfigure}
\begin{subfigure}{.31\linewidth}
\includegraphics[width=\linewidth]{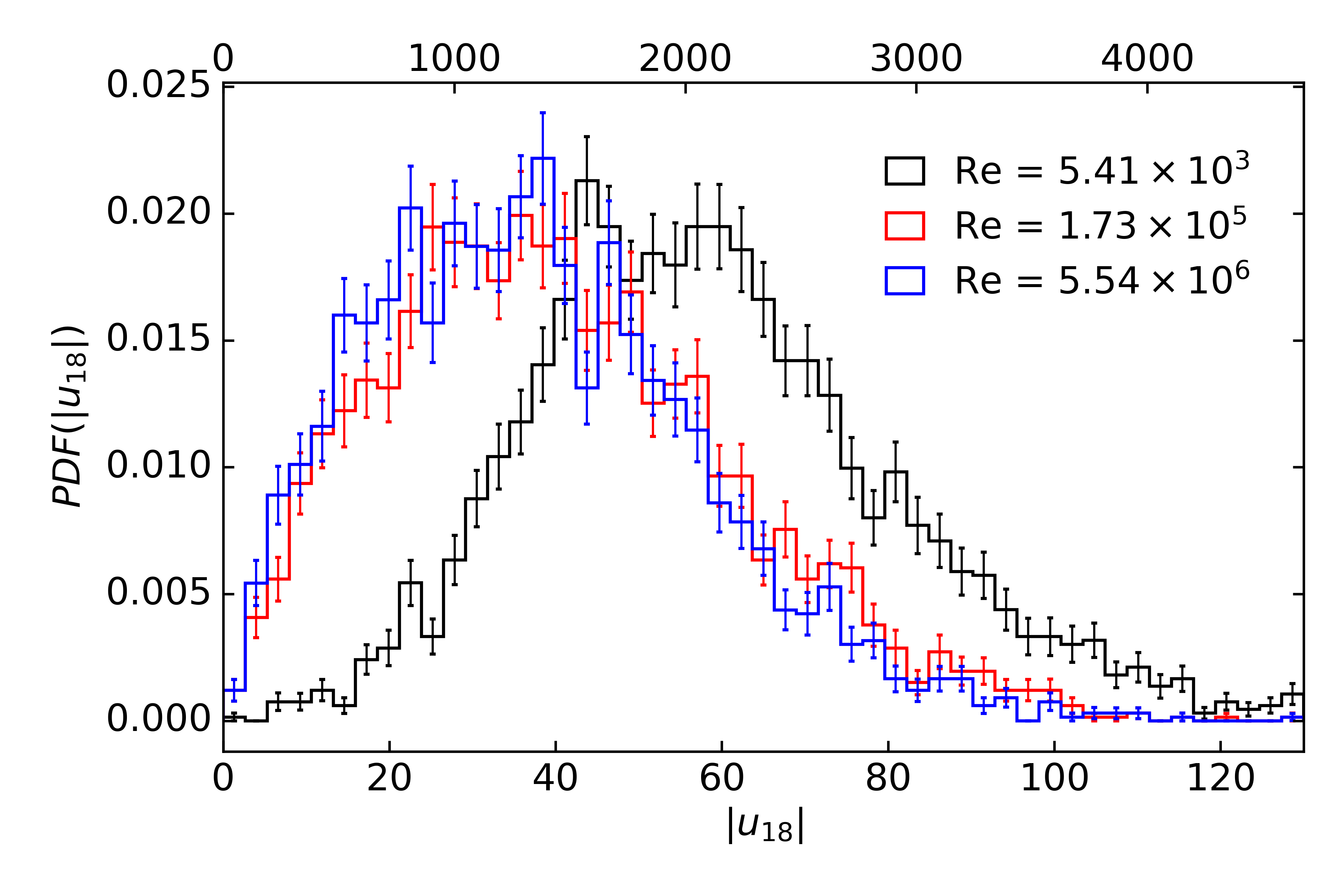}
\caption{``Burst'' initial state with noise scaling $\alpha = 0.$}
\end{subfigure}
\caption{Transition probability density function for the absolute values $|u_n|$ at a single fixed shellnumber $n=18$ and time $t_f=1$(a,b), $t_f=1.477\times 10^{-3}$ (c) in inertial units for Reynolds numbers spanning almost two decades. The bottom axis represents the inertial range units, while the top axis represents the SI units for the ABL parameters. All the errors are estimated as standard errors using the bootstrap method. \cite{suppmaterial} contains details how Reynolds number was varied, SM \S IV for the K41 initial condition, 
SM \S VI for the ``burst'' initial condition.} \label{pdf-stoch}
\end{figure*}

\textit{Spontaneous Stochasticity.} Spontaneous stochasticity can be given a precise meaning through a probability distribution on solutions of the governing stochastic differential equation. In the case of Eulerian spontaneous stochasticity triggered by thermal noise the corresponding equation is \eqref{FNS2}. Solution of this equation may be expressed in terms of a transition probability density $P_{Re, \Theta}(\bu_f, t_f|\bu_i, t_i)$ as follows
\be P_{Re, \Theta}\big(\bu_f,t_f)  = \int \mathcal{D}\bu_i \ P_{Re, \Theta}(\bu_f, t_f|\bu_i, t_i)  P_{Re, \Theta}(\bu_i, t_i), \lb{CKeq} \ee
where $P_{Re, \Theta}\big(\bu_{i/f} , t_{i/f} \big)$ is a probability distribution of velocity fields at the initial/final time. The transition probability $ P_{Re, \Theta}(\bu_f, t_f|\bu_i, t_i) $ satisfies the Fokker-Planck equation corresponding to Langevin equation \eqref{FNS2}, and it is parameterized by the non-dimensional numbers $Re$ and $\Theta$. In the limit of zero noise $\Theta \rightarrow 0$ with $Re$ fixed the transition probability becomes deterministic, that is, expressed as a delta-distribution on the unique solution of the limiting deterministic problem. The issue of uniqueness of solutions plays the central role in emergence of spontaneous stochasticity: see SM \S II \cite{suppmaterial} for more details. Note that the limiting equations for the fluctuating hydrodynamics of Landau-Lifschitz in the limit $\Theta \rightarrow 0$ is Navier-Stokes equations with the finite cut-off $\Lambda$, so that uniqueness of solutions to the Cauchy initial-value problem is then elementary and well-known. This physically necessary cut-off is crucial, since the uniqueness of Leray solutions of the continuum Navier-Stokes equations is a major open problem in pure mathematics \cite{fefferman2000existence}. 
%In contrast to some recent discussions \cite{palmer2014real}, we conclude that the non-uniqueness of Leray solutions is irrelevant to the issue of spontaneous stochasticity. 
However, if the zero-noise limit is taken together with $Re \rightarrow \infty$ and $\Lambda\propto Re\to\infty$, it leads to singular Euler dynamics with no unique solutions. Then the limiting transition probability may cease to become deterministic
\be P_{Re, \Theta}(\bu_f, t_f|\bu_i, t_i)  \xrightarrow[Re\rightarrow \infty]{\Theta \rightarrow 0} P_\infty (\bu_f, t_f|\bu_i, t_i). \label{doublelimit} \ee
If such a non-trivial limiting transition probability exists, the limit is called spontaneously stochastic and corresponds to stochastic behavior of a formally deterministic Euler system, which each realization of the limiting distribution satisfies in a weak sense \cite{bernard1998slow,falkovich2001particles,drivas2017lagrangian,mailybaev2016spontaneously,thalabard2020butterfly}.

Because the spontaneously stochastic limit is a double limit $Re^{-1}, \Theta \rightarrow 0$, there is no unique way to arrive at it. Furthermore, if the noise strength is taken to zero sufficiently fast, the limit becomes deterministic. In many cases of practical importance, such as turbulent flows past a grid or a cylinder, experimental evidence  \cite{sreenivasan1984scaling,cadot1997energy,sreenivasan1984scaling} points at a non-vanishing energy dissipation with the limiting dissipation rate satisfying a relation first proposed by Taylor \cite{taylor1935statistical}
$ \varepsilon = C {U^3}/{L}$,
where $C$ is a dimensionless constant. In this scenario we can control the macroscopic flow parameters $U$ (or $\varepsilon$) and $L$, while $\nu$, $T$, and $\rho$ are fixed material parameters of the fluid. These considerations motivate as a ``canonical limit'' the one obtained by fixing the ratio $U^3/L$, along with $\nu$, $T$, and $\rho$, while taking $Re \rightarrow \infty$. This leads us to define the second non-dimensional number as
$ \theta_\eta = {2k_B T}/{\rho \nu^{11/4}\varepsilon^{-1/4} }$
(see \cite{bandak2022dissipation}) and consider the limit
$ \Theta = Re^{-15/4} \theta_\eta \rightarrow 0 \label{double-limit} $
with $\theta_\eta$ held constant. This is the physically relevant continuum limit in which also $\Lambda\propto Re\to\infty,$ describing  fully-developed 3D hydrodynamic turbulence.

\textit{Numerical Verification of Eulerian Spontaneous Stochasticity.} In order to check that the limit \eqref{doublelimit} is indeed spontaneously stochastic we need to simulate Landau-Lifshitz equations at very high Reynolds numbers. State-of-the-art simulation can achieve only $Re \approx 500$ \cite{gallis2021turbulence,bell_nonaka_garcia_eyink_2022} for incompressible flows, so we use here the Sabra model \cite{l1998improved}, a simplified dynamical model of turbulent cascade that preserves many key features of Navier-Stokes equations \eqref{FNS2}, but discretizes length scales as $\ell_n = 2^{-n} \ L$ and keeps only one complex mode $u_n$ to represent velocity increments $\delta_l u \propto |u_n|$ at scale $\ell_n$. We also include a stochastic term to model thermal noise, and a deterministic forcing $f_n$ that acts only at large scales. In non-dimensionalized form this modified Sabra model is given by the stochastic ODEs:
\begin{multline} \label{stoch-sabra}
\frac{du_n}{dt} = i \Big( k_{n+1} u_{n+2} u_{n+1}^*- \frac{1}{2}k_{n} u_{n+1} u_{n-1}^* + \\ + \frac{1}{2}k_{n-1} u_{n-1} u_{n-2} \Big) - \frac{1}{Re} k_n^2 u_n + \sqrt{\Theta} k_n \xi_n(t) + F f_n,\cr
\qquad\qquad n=1,...,N
\end{multline}
where $k_n = 1/\ell_n$, covariance of the white noise $\langle\xi_n^*(t)\xi_m(t')\rangle=2 k_n^\alpha \delta_{nm} \delta(t-t)$, and the second non-dimensional number group is
$ \Theta =  Re^{-\beta} \theta_\eta$, $\beta = 3(\alpha + 2) /4$. Here we take the number of shells $N\propto \frac{3}{4}\log_2(Re),$ sufficient to resolve a few shells above the Kolmogorov wavenumber $k_\eta=1/\eta.$ The choice $\alpha=3$ in the noise covariance is dimensionally identical to 3D Landau-Lifschitz, with $\beta = 15/4$, and it produces also an energy spectrum $\propto k^2$  in the dissipation range, the same as for 3D fluids, but violates the shell-model fluctuation-dissipation relation . On the other hand the choice $\alpha=0$ preserves this relation although the equipartition energy spectrum in the dissipation range is $\propto k^{-1}$ rather than $\propto k^2$. Since it is impossible to match exactly all relevant properties of 3D Landau-Lifschitz equations with a single choice of $\alpha,$ we investigated both choices $\alpha=0$ and $\alpha=3$ and we find the overall results are insensitive to this choice. We would like to emphasize that for either choice of $\alpha$ the noise does not serve as a large scale forcing and %does not determine the range of forcing and dissipation. This is due to the fact that it 
in fact, together with the viscous damping, it becomes vanishingly small in the limit $Re\to\infty$. For more details on the numerical simulations, including the forcing $f_n$ used and the choice of $\alpha,$ see SM \S III-VI \cite{suppmaterial}. 

We study the Cauchy problem for \eqref{stoch-sabra} with two different deterministic but ``quasi-singular'' initial data that are not smooth uniformly in Reynolds number. It is convenient to study spontaneous stochasticity with such quasi-singular initial data since with large-scale initial data independent of $Re$ one would otherwise have to wait for singularities to form by finite-time blow-up \cite{mailybaev2016spontaneously}. The first is the Kolmogorov initial datum $u_{n} = -iA \varepsilon^{1/3} k_n^{-1/3},$ which is an exact stationary (but unstable) solution of the inviscid, deterministic Sabra model if suitable deterministic forces $f_n$ are added to the two lowest shells $n=1,2$ \cite{ditlevsen2010turbulence}. The other initial datum is a \lq\lq burst" state selected from the ensemble of turbulent steady states of the Sabra model at very high $Re$ (see SM \S V \cite{suppmaterial}). This particular initial datum has approximately a power-law form $u_n\propto k_n^{-h}$ in the inertial range, with H\"older exponent $h\simeq 0.258$; by construction, this is not intended to be the scaling of the statistical steady-state. Both of these initial data are quasi-singular with exponent $h<1,$ regularized only at very high-wavenumber either by the cut-off $N$ or by viscosity $\nu$. The numerical details of how $Re$ was varied differs for the two initial data:
see SM, \S IV for the K41 case, \S VI for the ``burst" case \cite{suppmaterial}.

The key statistical quantities which we calculate are the probability density functions (PDF's) of
local-in-scale variables, such as the absolute values of velocities at a fixed shellnumbers $n$, fixed time $t_f$, at an increasing sequence of Reynolds numbers. These reduced PDF’s are integrals over the transition probability densities in \eqref{CKeq}. Without external noise, these are delta-distributions; see SM, \S VII \cite{suppmaterial}.
Presented in Fig.~\ref{pdf-stoch} are plots of the PDF's for shell $n=18$ and time $t_f=1$(a,b), $t_f=1.477\times 10^{-3}$ (c), where panels (a),(b) are for the K41 initial datum with noise exponents $\alpha=0$ and $\alpha=3,$ respectively, and panel (c) is for the ``burst'' initial datum with $\alpha=0.$ As seen clearly, the PDFs converge with increasing $Re$ to non-delta distributions and therefore do not become deterministic. 
The direct effects of thermal fluctuations at this scale can be estimated from $\theta_{\eta}$ and the resulting RMS velocity fluctuations are 4-5 orders of magnitude smaller than the ones shown in Fig.~\ref{pdf-stoch}.
Thus, the universal statistics reflect spontaneous stochasticity not direct effects of thermal noise.
%
%This is true despite the fact that the noise term in \eqref{stoch-sabra} vanishes, rendering the limiting equations of motion deterministic. 
%
We have obtained similar results for the PDFs of other scale-local variables, e.g. energy flux $\Pi_n$ (see SM \S VIII \cite{suppmaterial}).

%These observations constitute our crucial numerical evidence for Eulerian spontaneous stochasticity triggered by thermal noise in the Sabra model and, presumably, for the Landau-Lifshitz equations. The two cases in panels (a) and (b) correspond to the same initial datum and the same limiting equations when $Re \rightarrow \infty$, but a different scale-by-scale approach towards it. Nevertheless, one can observe that the limiting probability distributions are the same. This illustrates an important property of spontaneous stochasticity, the independence of the limiting probability distributions from the type of regularization and the type of noise that triggers random perturbations. 

These observations constitute our crucial numerical evidence for Eulerian spontaneous stochasticity triggered by thermal noise in the Sabra model and, presumably, for the Landau-Lifshitz equations. The two cases in panels (a) and (b) correspond to the same initial datum and the same limiting equations when $Re \rightarrow \infty$, but a different scale-by-scale approach towards it. Nevertheless, the limiting probability distributions are the same, and independent of the type of regularization and the type of noise that triggers random perturbations.

\begin{figure}[]
\centering
\includegraphics[width=0.8\linewidth]{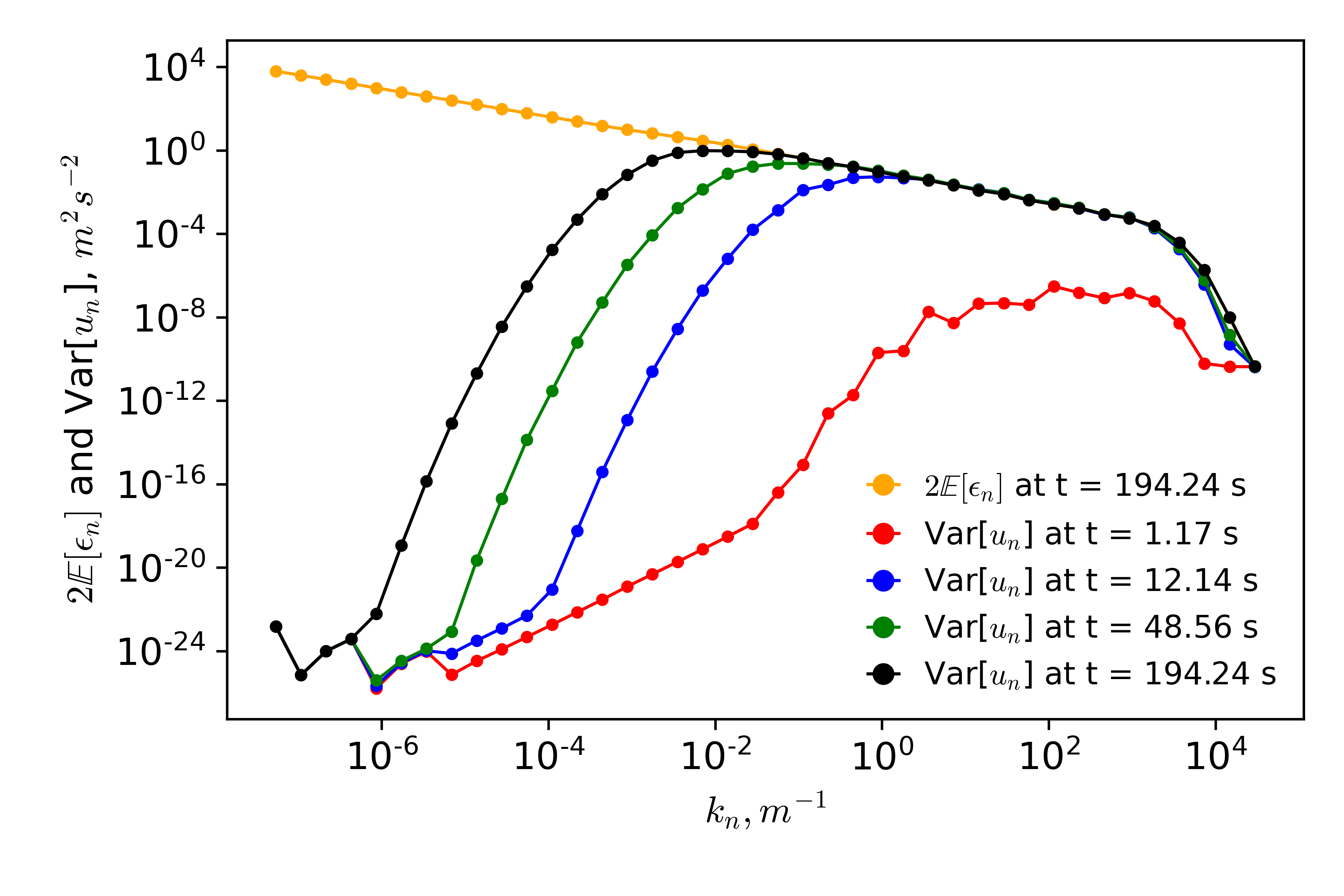}
\caption{Twice ensemble average energy $\mathbb{E}[\epsilon_n]$ (\textcolor{orange}{$\bullet$}) for $\epsilon_n=\frac{1}{2}|u_n|^2$ and velocity variances (\textcolor{blue}{$\bullet$},\textcolor{blue}{$\bullet$},\textcolor{green}{$\bullet$},\textcolor{black}{$\bullet$}) across the ensemble as a function of wavenumber in SI units for 4 increasing times. The smallest time in the variances plots the initial non-self-similar transient and the subsequent three times show the self-similar propagation of the stochastic wave towards large scales. $2\mathbb{E}[\epsilon_n]$ is almost unchanged in time and forms the envelope of the propagating wave.} \label{lorenz-plot}
\end{figure}

\textit{Inverse error cascade and stochastic wave.} 
What causes this unpredictability if the direct effects of thermal noise are too small?
The mechanism was first suggested by Lorenz: an {\it inverse cascade of error} \cite{lorenz1969predictability} that has since been extensively studied \cite{ngan2012middle, leith1972predictability, eyink1996turbulence, boffetta2001predictability}. Perhaps the simplest way to illustrate this mechanism is to look at the time-dependent variances ${\rm Var}[u_n]=\mathbb{E}[|u_n-\mathbb{E}[u_n]|^2]$ calculated across an ensemble of noise realizations with fixed initial datum. These are shown in Fig. \ref{lorenz-plot} for the K41 datum. Initially, variances at all scales exhibit diffusive linear growth in time, with higher rate at larger $k_n$. Next, modes become chaotic scale-by-scale, starting from high wavenumbers, and eventually the variance for a particular shell saturates when it reaches twice the average energy at that scale. In the early stage of development of the stochastic wave the total variance of the system $\mathrm{Var}(\bu)=\sum_n \mathrm{Var}(u_n)$ grows exponentially (see SM \S IX \cite{suppmaterial}), and this regime is fully consistent with work of Ruelle \cite{ruelle1979microscopic} on the effects of thermal noise in predictability of developed turbulence. However, when the stochastic front starts to propagate across the inertial range \cite{mailybaev2016spontaneously,biferale2018rayleigh} the system enters the spontaneously stochastic regime. In the case of a self-similar initial state $u_{0,n} \propto A k_n^{-h}$ the front is self-similar, located at length scale $\ell(t)=(At)^{1/(1-h)}$ with amplitude $\uit(t)=(At^h)^{1/(1-h)}$ at time $t.$ Plotted as $\mathrm{Var}(u_n)/\uit^2(t)$ versus $k_n\ell(t),$ the curves collapse for the three late times $t = 12.14, 48.56, 194.24 \ s$. For more details on the stages of the stochastic wave formation and propagation, see SM \S IX \cite{suppmaterial}. 
Furthermore, after the stochastic front passes some scale the statistics of Kolmogorov multipliers \cite{kolmogorov1962refinement} at that scale converges to the steady state distribution. Such {\it super-universality} has been observed before in \cite{mailybaev2016spontaneously}; see SM \S X \cite{suppmaterial}.  We draw attention to the striking resemblance of our Fig. \ref{lorenz-plot} to Figure 2 of Lorenz \cite{lorenz1969predictability}, which he obtained for 2-dimensional Euler equations using a turbulence closure model. For the analogous plot with the ``burst'' initial state, see SM \S XI \cite{suppmaterial}, where the same picture holds qualitatively, although there is no exact self-similarity.  The large spontaneous fluctuations illustrated in Fig.~\ref{pdf-stoch}(a) are thus due to effects of thermal noise in the dissipation range which are propagated up into the inertial range by nonlinear error cascade, and not due to the direct local effects of thermal noise.

\begin{figure}[]
\centering
\includegraphics[width=0.8\linewidth]{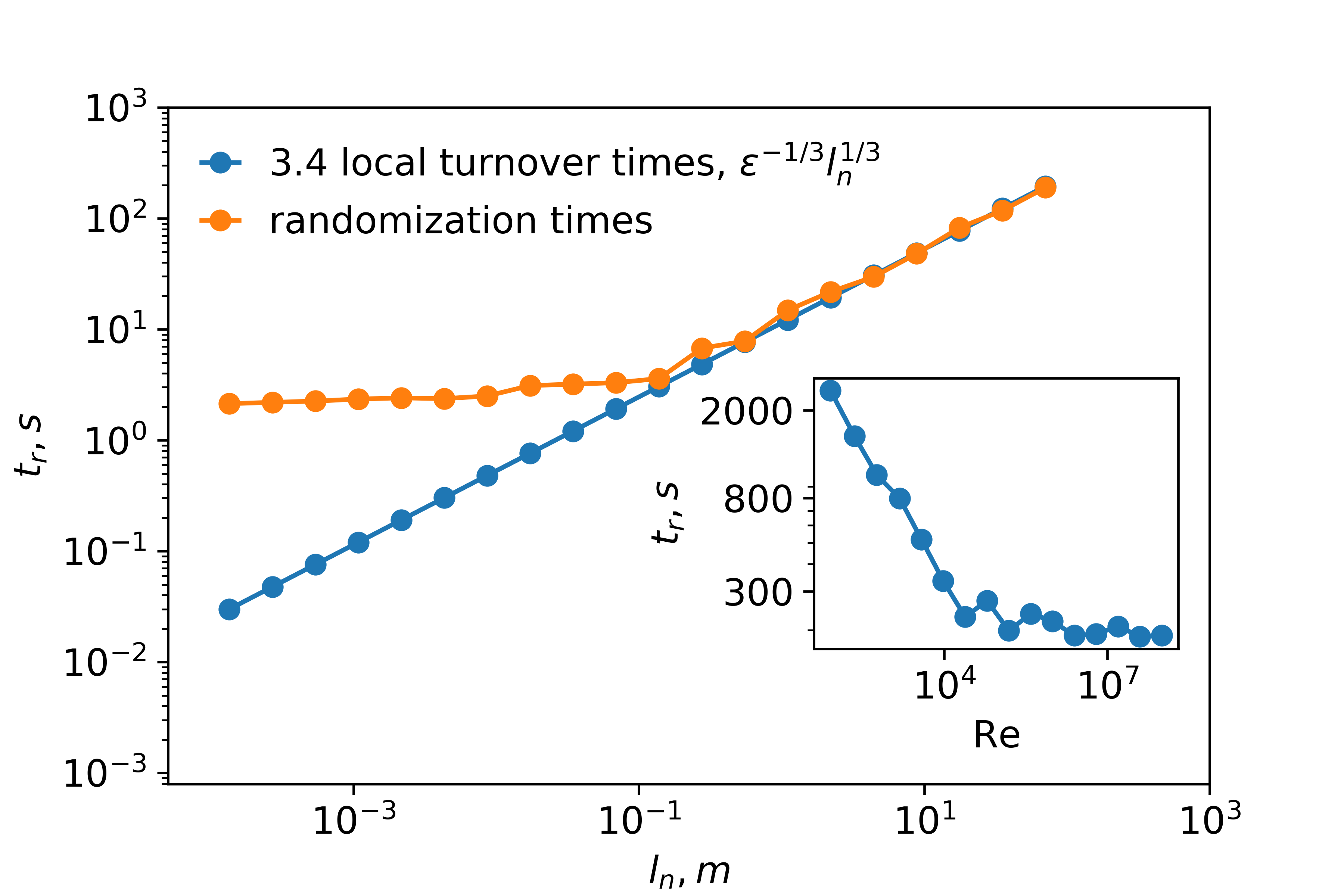}
\caption{Local randomization times $t_r(n)$ as a function of length-scale $\ell_n = 2^{-n}L$ for the K41 initial datum. $t_r(n)$ is defined as the time in which the $n$th shell's variance reaches the ensemble average energy  $\mathbb{E}[\epsilon_n]$. The inset plot depicts $t_r(18)$ as a function of Reynolds number.} \label{random-times}
\end{figure}

An important feature of this ``inverse error cascade'' is that in  the inertial range the universal statistical distributions are achieved at each length-scale $\ell$ in a time which is a constant multiple of the eddy-turnover time $\tau_\ell=\ell/u_\ell$, indifferent to the noise magnitude. This should be contrasted with a predictability horizon in conventional chaotic systems, which is dependent on the noise strength \cite{lorenz1969predictability,palmer2014real, palmer2019stochastic, thalabard2020butterfly}. 
%In the words of Lorenz \cite{lorenz1969predictability} ``two states of the system differing initially by a small `observational error' will evolve into two states differing as greatly as randomly chosen states of the system within a finite time interval, which cannot be lengthened by reducing the amplitude of the initial error.'' 
To illustrate this point, Fig. \ref{random-times} shows the randomization time $t_r(n)$, defined as the time when the $n$th shell's variance reaches its ensemble average energy, plotted versus index $n$. As is clear from the figure, the randomization times above the lengthscale of $10 \ \mathrm{cm}$ for the flow parameters of the ABL are given by $t_r(n)=3.4 \varepsilon^{-1/3} \ell_n^{2/3}$. Therefore we conclude that the length scales of about the size of a coffee mug and above in 3D ABL turbulence behave in a spontaneously stochastic fashion. In the SM \S XII \cite{suppmaterial} we provide a theoretical estimate on dimensional grounds of that length scale as a function of $Re$ and $\Theta$. Crucially, we observe that $t_r(n)$ approaches the asymptotic value $\propto \varepsilon^{-1/3} \ell_n^{2/3}$ for any shell $n$ in the limit $Re\to\infty:$ see inset of Fig.~\ref{random-times}. Thus {\it all} scales are spontaneously stochastic in that idealized limit.

\textit{Discussion.} 
%It is important to emphasize that {\it any} noise source of not too small amplitude for $Re\gg 1$ will give rise to the same spontaneous statistics. As discussed in \cite{eyink2020renormalization}, it is possible for the solution to remain deterministic if the noise vanishes too quickly as $Re\to\infty.$ A crucial contribution of the present work is to demonstrate that inevitable thermal noise does {\it not} vanish too rapidly as $Re\to\infty$ and suffices to trigger spontaneous stochasticity. 
%\textcolor{blue}{It is important to emphasize that the spontaneous large-scale statistics are universal, arising from {\it any} small-scale noise source whose amplitude becomes negligible with respect to the deterministic equation more slowly than $\exp(-1/\sqrt{Re})$ as $Re \rightarrow\infty$ (see SM \S XII). 
%Our work demonstrates that even the inevitably present thermal noise satisfies this criterion and is sufficient to trigger spontaneous stochasticity.}
%
\textcolor{black}{It is important to emphasize our finding that the spontaneous large-scale statistics are universal with respect to the small-scale noise that triggers them, as long as the noise amplitude becomes negligible with respect to the deterministic equation more slowly than some $Re$-dependent threshold. On dimensional grounds we estimate this threshold to be $\sim \exp(-\sqrt{Re})$ as $Re \rightarrow\infty$ (see SM \S XII \cite{suppmaterial}). Even the inevitably present molecular noise satisfies this criterion and our simulations suggest that it is sufficient to trigger spontaneous stochasticity.}
%A crucial contribution of the present work is to demonstrate that inevitable thermal noise does {\it not} vanish too rapidly as $Re\to\infty$ and suffices to trigger spontaneous stochasticity. 
%In the real atmosphere and ocean there will be other noise sources, e.g. the proverbial flap of a butterfly wing or even swimming motions of micro-organisms, generating hydrodynamic perturbations larger than those of molecular agitation. What is crucial about thermal noise is that it is {\it unavoidable}, because it is a direct consequence of the local thermodynamic equilibrium which enables a hydrodynamic description to exist at all. 
In one turnover time of the largest 3D turbulent eddies the unknown molecular motions will impact the evolution, rendering only statistical predictions possible.
%\textcolor{red}{Our work refines the pioneering work of Lorenz  \cite{lorenz1969predictability} by showing that universal statistics are attained in the limit $Re\to\infty$ and give precise meaning to his claim that ``formally deterministic fluid systems'' may possess stochastic solutions. See SM, \S II.}

Our work has implications for turbulence across multiple scales. %(see SM \S XII \cite{suppmaterial}). 
For climate models, even if the projected goal of 1 km horizontal resolution in the next decade is achieved \cite{palmer2019stochastic},  such refined resolution will not obviate the need for stochastic models \cite{hasselmann1976stochastic,palmer2014real,palmer2019stochastic,lovejoy2022future}.  For the dynamics of galaxy formation, it has already been shown that microscale chaos and stochasticity lead to large variations in star-formation histories and distribution of stellar mass \cite{keller2019chaos}, and our results suggest that these effects may be even more severe than currently thought.  At the large scale of hydrodynamic simulations of cosmological galaxy formation, the sensitivity of simulations to minute perturbations has also been examined with regard to chaotic dynamics \cite{genel2019quantification}, and would be expected to be amplified further by the results we have discussed \cite{neyrinck2022boundaries}. Closer to home, there have been recent efforts to reconstruct best-fit individual solution trajectories of Navier-Stokes equations using variational data assimilation techniques \cite{buzzicotti2020synchronizing, zaki2021limited}.  It is already recognized that these reconstruction problems are highly ill-conditioned due to chaotic dynamics.  The inclusion of spontaneous stochasticity into this program poses even more severe limitations and implies that a well-posed problem is instead the reconstruction of statistical ensembles of solutions \cite{mons2021ensemble,biferale2017optimal}.  These examples show that there are many potential ramifications of spontaneous stochasticity in turbulence and related phenomena.  It will be important to determine if our findings, based admittedly on a shell model, are valid beyond the necessary simplifications entailed in our work.

This work was partially supported by the Simons Foundation through Targeted Grant \lq\lq Revisiting the Turbulence Problem Using Statistical Mechanics" (Grant Nos. 663054 (G.E.) and 662985(N.G.)). A.A.M. was supported by the CNPq grant 308721/2021-7 and FAPERJ grant E-26/201.054/2022.

\clearpage
\onecolumngrid

\tocless\section{Supplemental Material}

\setcounter{tocdepth}{3}
\setcounter{secnumdepth}{3}

\tableofcontents

\section{Landau-Lifschitz Fluctuating Hydrodynamics and the Role of the UV Cut-Off $\bLambda$} Here we review briefly the equations of
Landau-Lifschitz  fluctuating hydrodynamics and the role of the wavenumber cut-off $\Lambda.$ We explain also why taking the limit $\Lambda\to\infty$
at fixed $Re,$ $\Theta$ to obtain a continuum SPDE is physically irrelevant to the understanding of 3D fluid turbulence.

The fluctuating hydrodynamics equations for an incompressible fluid are given in dimensional form by:
\be \partial_t\bu + (\bu\cdot\grad)\bu = -\grad p + \nu\triangle \bu + \sqrt{\frac{2\nu k_B T}{\rho}} \grad\cdot \boeta, \qquad \grad\bdot\bu =0 \lb{FNS2} \ee
where $\nu$ is kinematic viscosity, $T$ is absolute temperature, $k_B$ is Boltzmann's constant, $\rho$ is mass density, and the fluctuating stress is modeled as a Gaussian random matrix field $\boeta$ with mean zero and covariance
\begin{eqnarray}
	\langle \xi_{ij}(\bx,t) \xi_{kl}(\bx',t')\rangle& = &
	\left(\delta_{ik}\delta_{jl}+\delta_{il}\delta_{jk} -\frac{2}{3}\delta_{ij}\delta_{kl}\right)
	\delta^3_\Lambda(\bx-\bx')\delta(t-t') \lb{FDR}.
\end{eqnarray}
For example, see \cite{forster1977large, bandak2022dissipation}.  Here we consider these equations for simplicity in a periodic box and note that $\delta^3_\Lambda(\br)$ in \eqref{FDR} is a ``cut-off delta-function'', defined as \cite{zubarev1983statistical}
\be \delta^3_\Lambda(\bx-\bx')=\frac{1}{V} \sum_{|\bk|<\Lambda} e^{i\bk\bdot\br}. \ee
In the microscopic derivation of these equations using Zwanzig-Mori techniques, e.g. see \cite{zubarev1983statistical,morozov1984langevin,espanol2009microscopic}, the viscosity is given by Green-Kubo-type formulas which show that $\nu=\nu_\Lambda$ must be cut-off dependent. This $\Lambda-$dependence of the effective viscosity has been studied also by renormalization-group methods for a quiescent fluid in thermal equilibrium \cite{forster1977large}. In practical computations the values of $\nu_\Lambda$ are often fixed by comparison with molecular dynamics simulations, as in \cite{donev2014low}, Appendix C.

The choice of the high-wavenumber cut-off $\Lambda$ in the Landau-Lifschitz equations \eqref{FNS2} is somewhat arbitrary, but subject to important constraints. The microscopic derivations  \cite{zubarev1983statistical,morozov1984langevin,espanol2009microscopic} show that $\Lambda$ must be chosen somewhat smaller than the microscopic wavenumber $1/\lambda_{micr},$ where $\lambda_{micr}=\max\{\lambda_{mfp},\lambda_{intp}\}$ with $\lambda_{mfp}$ the mean-free-path length and $\lambda_{intp} =n^{-1/3}$ the interparticle spacing (where $n=\rho/m$ is the number of molecules of mass $m$ per volume). Note that $\lambda_{micr}=\lambda_{mfp}\gg \lambda_{intp}$ in a low density gas and $\lambda_{micr}=\lambda_{intp}\gtrsim \lambda_{mfp}$ in a liquid. Empirically, the model \eqref{FNS2} is found to be accurate with $\Lambda$ just moderately smaller than $1/\lambda_{micr}$ (e.g. see \cite{donev2011enhancement}). On the other hand, $\Lambda$ must not be chosen too small, so that the only velocity fluctuations which are eliminated by coarse-graining correspond to local thermodynamic equilibrium fluctuations. In a turbulent flow, this implies that, at least, $\Lambda\gtrsim 1/\eta,$ with $\eta$ the Kolmogorov length, so that the eliminated wavenumber modes are not subject to turbulent fluctuations. However, as discussed at some length in our previous work \cite{bandak2022dissipation}, intermittency effects can lead to turbulent effects propagating to sub-Kolmogorov lengths much smaller than $\eta.$ In that case, it would be safer to choose $\Lambda$ just a factor of a few smaller than $1/\lambda_{micr}.$ Note that the integral length $L$ and $\eta$ are related through the scaling $L/\eta\simeq Re^{3/4}$ using Taylor's relation $\varepsilon\propto U^3/L$.  $L$ and $\lambda_{mfp}$ are related through $L/\lambda_{mfp}\simeq Re/Ma$ using the kinetic theory estimate $\nu=\lambda_{mfp}c$ where $c$ is the sound speed.  In dimensionless units in terms of the outer scales, it therefore follows that the cut-off wavenumber at the Kolmogorov-scale is $\hat{\Lambda}:=\Lambda L\sim Re^{3/4}$ whereas the cutoff at the molecular scale is $\hat{\Lambda}\sim Re/Ma.$ It should be emphasized that even the latter cut-off may not be sufficient to guarantee validity of fluctuating hydrodynamics. As discussed in \cite{bandak2022dissipation}, extremely large, rare events could produce turbulent fluctuations at the length scale $\lambda_{micr}.$ If such extreme events occur, then the Landau-Lifschitz equations \eqref{FNS2} would break down, at least locally in space-time.

The well-known renormalization group analysis of Forster et al. \cite{forster1977large} suggests that there is a UV strong-coupling regime in model (\ref{FNS2}) in 3D, because the ``Reynolds number''  $Re_\theta=u_\Lambda^\theta/\nu_\Lambda\Lambda$ of thermal velocity fluctuations $u^\theta_\Lambda=(k_BT \Lambda^3/\rho)^{1/2}$ increases with $\Lambda$ for fixed $Re$ (even for $Re\ll 1!$) Although the equations (\ref{FNS2}) are {\it not} continuum stochastic partial-differential equations, because of the cut-off $\Lambda,$ the mathematical methods of stochastic partial differential equations (SPDE's) developed to study the formal $\Lambda\to\infty$ limit might be relevant to describe the UV strong-coupling regime at finite $Re$, if the latter occurred in a physical range of wavenumbers.  However, simple estimates \cite{bandak2022dissipation} show that the nonlinear coupling becomes strong in \eqref{FNS2} only for $\Lambda^{-1}$ of order the mean-free-path length, or smaller, where any hydrodynamic description breaks down! This is in contrast to the situation for similar stochastic models in less than two dimensions, such as 1D KPZ, where the RG analysis of \cite{forster1977large} showed that the strong-coupling regime occurs instead in a physical regime {\it at low wavenumbers}. Thus, the methods of stochastic PDE's to study the limit $\Lambda\to\infty$ for KPZ \cite{hairer2013solving} are physically very relevant, since that limit corresponds to the nontrivial IR-attractive RG fixed point which governs KPZ scaling  at low wavenumbers \cite{corwin2015renormalization}. In other words, the infrared limit $k\ll \Lambda$ for a microscopic stochastic growth model with physical UV cutoff $\Lambda$ corresponds mathematically, by a simple rescaling, to holding wavenumber $k$ fixed and taking $\Lambda\to \infty.$

The latter limit should be clearly distinguished from that considered in our work. Here we keep the dimensional UV cutoff $\Lambda$ fixed at some value between $1/\eta$ and $1/\lambda_{micr},$ while taking $Re=UL/\nu\to \infty.$ In this limit, the outer-scaled cutoff $\hat{\Lambda}$ lies between $Re^{3/4}$ and $Re/Ma$ and thus $\hat{\Lambda}\to\infty,$ similar to KPZ. Simultaneously, however, the dimensionless viscosity $\hat{\nu}=1/Re$ and thermal noise $\Theta=Re^{-15/4}\theta_\eta$ both {\it vanish}, so that the solution of the Cauchy problem for \eqref{FNS2} with initial data $\bu_0$ is described in this limit by a statistical ensemble of solutions to deterministic Euler equations, each with initial datum $\bu_0$ \cite{eyink2023infinite}. The infinite $Re$ limit is governed not by an SPDE but instead by a deterministic PDE, the Euler equations, with spontaneously stochastic solutions.

\section{Spontaneous Stochasticity and Non-uniqueness of Solutions}
\textcolor{black}{The concept of spontaneous stochasticity was first introduced in the seminal paper \cite{bernard1998slow} that studied the Kraichnan model of passive scalar advection (see also \cite{falkovich2001particles}). This work showed that trajectories of Lagrangian particles advected by a rough velocity field remain stochastic even in the limit of zero noise, a phenomenon now known as Lagrangian spontaneous stochasticity. The limiting distribution is over the set of non-unique particle trajectories and is responsible for the anomalous dissipation of scalars advected by turbulence \cite{bernard1998slow,drivas2017lagrangian} and the emergence of universal particle statistics in finite time. This is in sharp contrast to chaos, where universality is associated with infinite times. Eulerian spontaneous stochasticity is a related phenomenon with the limiting distribution over the non-unique velocity fields solving the Euler equation \cite{mailybaev2016spontaneously,thalabard2020butterfly}. Recently, by PDE methods of convex integration it has been shown that there is at least a dense set of ``rough'' initial data for which infinitely many solutions of the Cauchy problem for the Euler equations exist, even if one imposes the conditions that the Euler solutions are locally dissipative and H\"older continuous in spacetime \cite{de2010admissibility,de2020non,daneri2021non} as required for the energy cascade by Onsager's theorem \cite{onsager1949statistical,eyink2006onsager}. Such non-uniqueness, as illustrated by the more tractable example of Lagrangian spontaneous stochasticity, may lead to Eulerian spontaneous stochasticity,}
\textcolor{black}{as has been recently verified rigorously in some simple toy models
	\cite{mailybaev2021spontaneously,mailybaev2022spontaneous}.}

\textcolor{black}{
	It is clear in retrospect that Lorenz in his seminal work \cite{lorenz1969predictability} anticipated several key components of the modern concept of spontaneous stochasticity.  Lorenz understood the essential difference between now-standard chaos and spontaneous stochasticity, stating that the predictability horizon of the latter would not increase with decreasing noise. Lorenz recognized also that this fundamental property would lead to essentially stochastic dynamics. In his own words from \cite{lorenz1969predictability}, ``certain formally deterministic fluid systems which possess many scales of motion are observationally indistinguishable from indeterministic systems''. What is added here is the clear understanding that spontaneous stochasticity is a phenomenon that occurs in a singular limit such as $Re\to\infty$ \cite{bernard1998slow}. In this limit, Lyapunov exponents are not merely positive, as in standard chaos, but in fact diverge to positive infinity. In Eulerian stochasticity the divergence happens at small scales, while in the Lagrangian case divergence happens at rough points of the fluid flow. Ultimately this is the root cause of the qualitative difference in predictability horizon of chaotic and spontaneously stochastic systems. We furthermore stress the importance of the insight of  \cite{bernard1998slow} that connects spontaneous stochasticity with non-unique solutions of a limiting singular initial-value problem. It is precisely the non-uniqueness of singular solutions of the limiting ideal dynamics which permits the randomness from vanishingly small stochastic perturbations to persist and to yield robust, universal statistics in a finite time, independent of the noise source. \textcolor{black}{The ``formally deterministic fluid system'' envisioned by Lorenz to have stochastic solutions 
		must be the ideal Euler equation, and not the viscous Navier-Stokes equation. The latter equation at finite Reynolds number exhibits spontaneous statistics only if it is subject to tiny random perturbations, either in the dynamics or in the initial data, which vanish in the limit $Re\to\infty$. The stochastic behavior of solutions of the deterministic Euler equations
		requires not merely ``many scales of motion'', but in fact infinitely many infinitesimally-small scales in the asymptotic high-$Re$ limit.}
	We believe that identification of the ``real butterfly effect'' \cite{palmer2014real} of Lorenz with Eulerian spontaneous stochasticity clarifies the mathematical foundations of this effect and should accelerate further progress. 
}

\section{Shell Model and Numerical Methods}
\textcolor{black}{The recent state-of-the-art simulations of Bell \textit{et al.} \cite{bell_nonaka_garcia_eyink_2022} could attain only $Re=554$, which is orders of magnitude below the value in the atmospheric boundary layer, for example. Likewise, Gallis et al. \cite{gallis2021turbulence} using the molecular gas dynamics method of Direct Simulation Monte Carlo (DSMC) could achieve only $Re=500$ for nearly incompressible flow and $Re=2000$ for supersonic flow. While this latter paper did observe effects of thermal noise on large scales, their Reynolds numbers were not high enough to verify the convergence of statistics which is the signature of spontaneous stochasticity. Therefore to verify our claims it was necessary to use the Sabra shell model supplemented with an appropriate thermal noise term.}

Dynamics of velocities $u_n$ in the noisy Sabra model in its nondimensional form is governed by the following system of stochastic ODEs (equation [9] in the main text):
\begin{multline} \label{stoch-sabra}
	\frac{du_n}{dt} =  V_n[u,u^*] + \sqrt{\Theta} k_n \xi_n(t) = B_n[u, u^*] +f_n - \frac{1}{Re} k_n^2 u_n + \sqrt{\Theta} k_n \xi_n(t) = \\
	= i \Big( k_{n+1} u_{n+2} u_{n+1}^*- \frac{1}{2}k_{n} u_{n+1} u_{n-1}^* + \frac{1}{2}k_{n-1} u_{n-1} u_{n-2} \Big) - \frac{1}{Re} k_n^2 u_n + \sqrt{\Theta} k_n \xi_n(t),
\end{multline}
where $k_n = 2^n/L$ are wavenumbers, $Re=UL/\nu$ is Reynolds number, $f_n$ represent a deterministic external forcing to drive turbulence, and $\xi_n$ are white noises that model thermal fluctuations, so that $\Theta=2\nu k_BT/\rho L^{1+\alpha}U^3$ is the second dimensionless number due to the presence of the thermal noise term. The latter depends upon the scaling power $\alpha$ of the covariance of white noise $\langle\xi_n^*(t)\xi_m(t')\rangle=2\delta_{nm} k_n^\alpha \delta(t-t)$, which is an important degree of freedom at our disposal. As noted in the main text, we may write alternatively $\Theta=Re^{-\beta}\theta_\eta$ with $\theta_\eta = 2k_BT/\rho\nu^{(3\alpha+2)/4}\varepsilon^{(2-\alpha)/4}$ and with $\rho$ assigned dimensions $({\rm mass})/({\rm length})^\alpha$. In our work we studied two different values, $\alpha = 0$ and $\alpha=3$. The noise spectrum given by $\alpha = 0$ leads to high-wavenumber modes that are in equipartition and form a thermal bath. This choice implies that there is detailed balance in the stochastic dynamics of small-scale modes. This can be seen from the Fokker-Plank operator ${\mathcal L}$
\be {\mathcal L} P=  \sum_{n=0}^N \Bigg( -\frac{\partial}{\partial u_n} \Big(V_n[u,u^*]P\Big)  - \frac{\partial}{\partial u_n^*} \Big(V^*_n[u,u^*]P\Big) + 2 \Theta k_n^{2+\alpha} \frac{\partial^2}{\partial u_n\partial u_n^*}P\Big) \ee
of stochastic dynamics \eqref{stoch-sabra} with $f_n=0$, which is self-adjoint with respect to the Gaussian equilibrium measure $P_G[u, u^*]= \mathrm{exp} \big\{ - \frac{1}{Re\,\Theta} \sum_{n=0}^N \epsilon_n \big\}, \epsilon_n = \frac{1}{2}|u_n|^2$ for $\alpha=0$. See \cite{bandak2022dissipation}, Appendix A. In molecular fluids, the validity of the hydrodynamic approximation is predicated on the assumption of local thermodynamic equilibrium. This is indeed equivalent to high-wavenumber modes remaining in thermal equilibrium, which is why the case $\alpha = 0$ is a good choice to represent this feature of molecular fluids.

\begin{figure}
	\centering
	\includegraphics[width=0.6\linewidth]{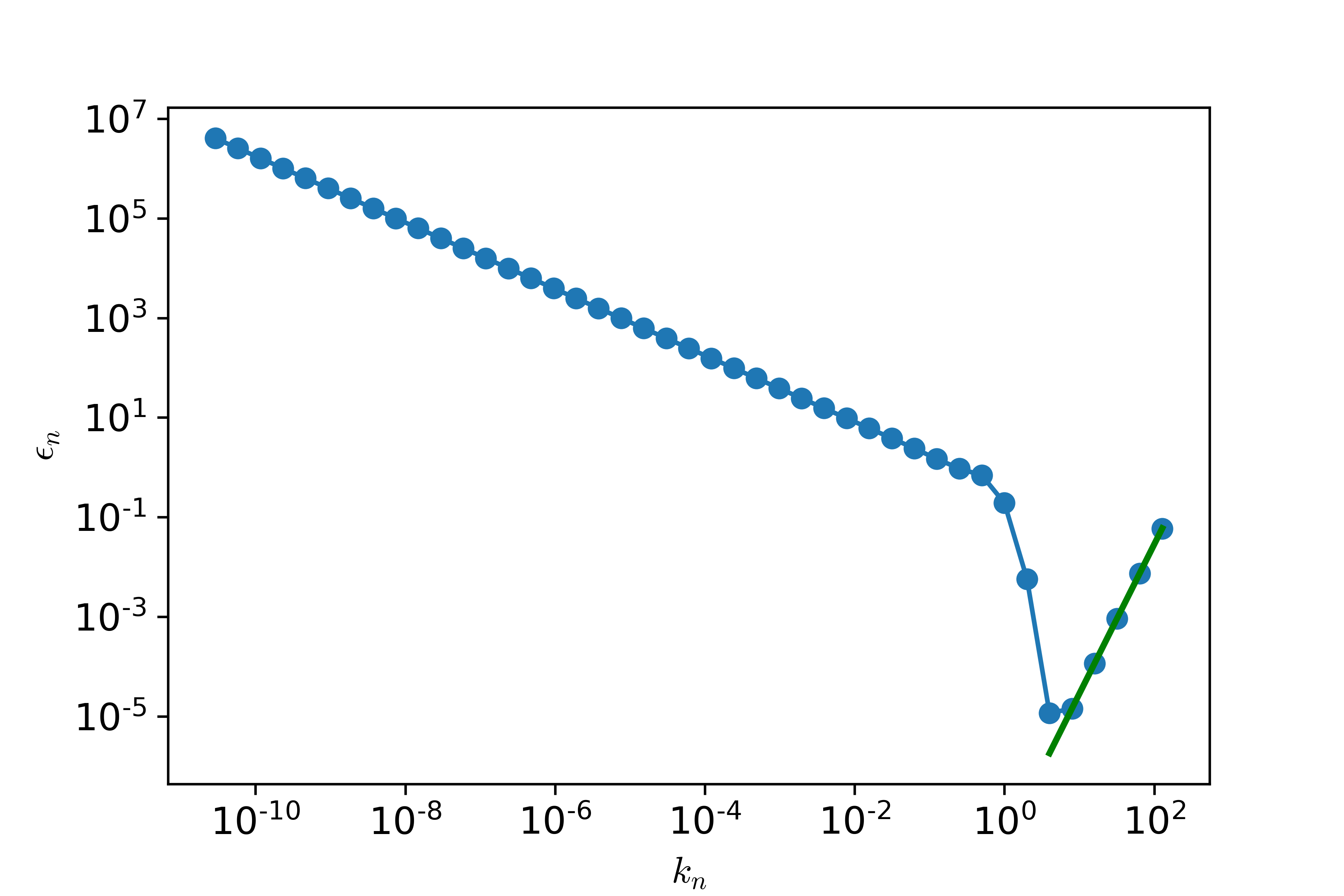}
	\caption{Energy ensemble average as a function of wavenumber at time $t = 1$ in dissipation units for the K41 initial datum in a simulation with noise exponent $\alpha=3$. The modal energy at high-wavenumbers is fit to power-law scaling, with fit depicted in green. The power-law fit has the power of 3.00}. \label{k2-spectrum}
\end{figure}

In contrast, the noise spectrum given by $\alpha = 3$ leads to breaking of detailed balance and therefore the high-wavenumber modes are not in equipartition. However, in this case the energy spectrum $E_n:=\epsilon_n/k_n$ scales as $k^2_n,$ as is expected in 3D fluids, and thus closer to 3D turbulence from this perspective. The $k_n^2-$scaling can be expected from the balance of the viscous and noise term in \eqref{stoch-sabra}, neglecting the nonlinear interactions. To numerically verify this expectation, we ran a simulation with the K41 initial data $u_n=-iA\epsilon^{1/3}k_n^{-1/3}.$ Figure \ref{k2-spectrum} shows the modal energy $\epsilon_n$ averaged over different noise realizations as a function of wavenumber at time $t = 1$ in dissipation units. The modal energy at high-wavenumbers is fit to a power-law scaling, with fit depicted in green. The power-law fit has the power of 3.00, and thus the spectrum $E_n\propto k_n^2,$ as expected. \textcolor{black}{Since it is impossible to match exactly all of the relevant properties of 3D Landau-Lifschitz equations with a single choice of $\alpha,$ we investigated both choices $\alpha=0$ and $\alpha=3$ and the overall results are insensitive to this choice.}

Stochastic integration of \eqref{stoch-sabra} has been performed using a slaved Taylor-Ito scheme from \cite{kloeden2011exponential} which is nominally $\frac{3}{2}$-order but, in practice, 2nd-order. The details of the derivation of the scheme for stochastic Sabra model may be found in reference \cite{bandak2022dissipation}, and here we just state the update rule for complex velocity $u_n$
\bea
&& u_n(t_{k+1})= e^{-\nu k_n^2 \Delta t}\Bigg\{u_n(t_k) +\Delta t[B_n(t_k,u(t_k))+f_n(t_k)] +\frac{1}{2} (\Delta t)^2 (\nu k_n^2[B_n(t_k,u(t_k))+  f_n(t_k)]+ \dot{f}_n(t_k)) + \cr
&& + \frac{1}{2}(\Delta t)^2 \sum_m \Big[a_m \frac{\partial B_n}{\partial u_m}(t_k,u(t_k))+ a_m^* \frac{\partial B_n}{\partial u_m^*}(t_k,u(t_k))  +2b_m^2 \frac{\partial^2 B_n}{\partial u_m\partial u_m^*}(t_k,u(t_k))\Big] + \cr
&& \sum_m b_m\Big[\Delta Z_m(t_k) \frac{\partial B_n}{\partial u_m}(t_k,u(t_k)) +\Delta Z_m^*(t_k) \frac{\partial B_n}{\partial u_m^*}(t_k,u(t_k))\Big] + b_n [(1 +\nu k_n^2 \Delta t)\Delta W_n(t_k) - \nu k_n^2\Delta Z_n(t_k) ]\Bigg\} \cr
&& \lb{LR04}
\eea
\be B_n(u)=V_n[u,u^*], \quad a_n = B_n +f_n, \quad b_n=\sqrt{\Theta} k_n, \ee
where $f_n$ is forcing at shell $n$.

\section{Stationary Kolmogorov self-similar solution, Reynolds number and time-inertial units} A few words are required to explain how to define a ``Reynolds number'' for the K41 solution $u_n=-iA\epsilon^{1/3}k_n^{-1/3},$ which has no natural large scale $L.$ Note that we can truncate this self-similar state at any arbitrarily large scale, the dynamical effect of which is negligible at short enough times. As a suitable proxy for $L$, we instead introduce an ``observation time'' $t_o$, which we will employ for non-dimensionalization. To that end we state dimensions of all the basic dimensionful quantities
\begin{gather*}
	[t] = [time] ,  \, [k_n] = [length]^{-1}  , \, [u] = [length] [time]^{-1} , [f] = [length][time]^{-2} , \cr
	[\nu] =  [length]^{2} [time]^{-1}, \, \Big[\frac{k_B T}{\rho} \Big] = [length]^2 [time]^{-2}, \cr
	[\epsilon]  = [length]^2[time]^{-3}, \, [\eta] = [time]^{-1/2}.
\end{gather*}
This leads us to introduce a length-scale $l_o = \epsilon^{1/2} t_o^{3/2}$, which we note from Fig.~4 in the main text is the length scale reached by the spontaneously stochastic
wave at time $t_o.$  The corresponding velocity is $u_o=(\epsilon l_o)^{1/3} = (\epsilon t_o)^{1/2}.$ We then non-dimensionalize the equations with these scales to obtain the Sabra model dynamics in ``time inertial units'' as
\be  \frac{d u_n}{d t} = B_n [u_i] - \frac{1}{Re_o}  k_n^2 u_n + f_n + \sqrt{\Theta_o} k_n \xi_n(t), \ee
upon introduction of dimensionless parameters $Re_o = \frac{u_o l_o}{\nu}=\frac{\epsilon t_o^2}{\nu}$ and $\Theta_o = \frac{2k_B T }{\rho u_o^3l_o^{1+\alpha}} = Re_o^{-3(2+\alpha)/4} \theta_\eta.$ In this manner, the limit of large Reynolds number $Re_o\gg 1$ over a unit time interval is achieved by simulating the model at $Re_o=1,$
$\Theta_o=\theta_\eta$ for a long ``observation time'' $t_o$ and then non-dimensionalizing as discussed above. The Reynolds numbers 
$Re=6.25\times 10^6,$ $15.75\times 10^6$, $39.68\times 10^6$, $100\times 10^6$ reported in Figure 1 were achieved with 
$t_o=2500,$ $3968,$ $6299,$ $10000.$ 

\begin{figure}
	\centering
	\includegraphics[width=0.45\linewidth]{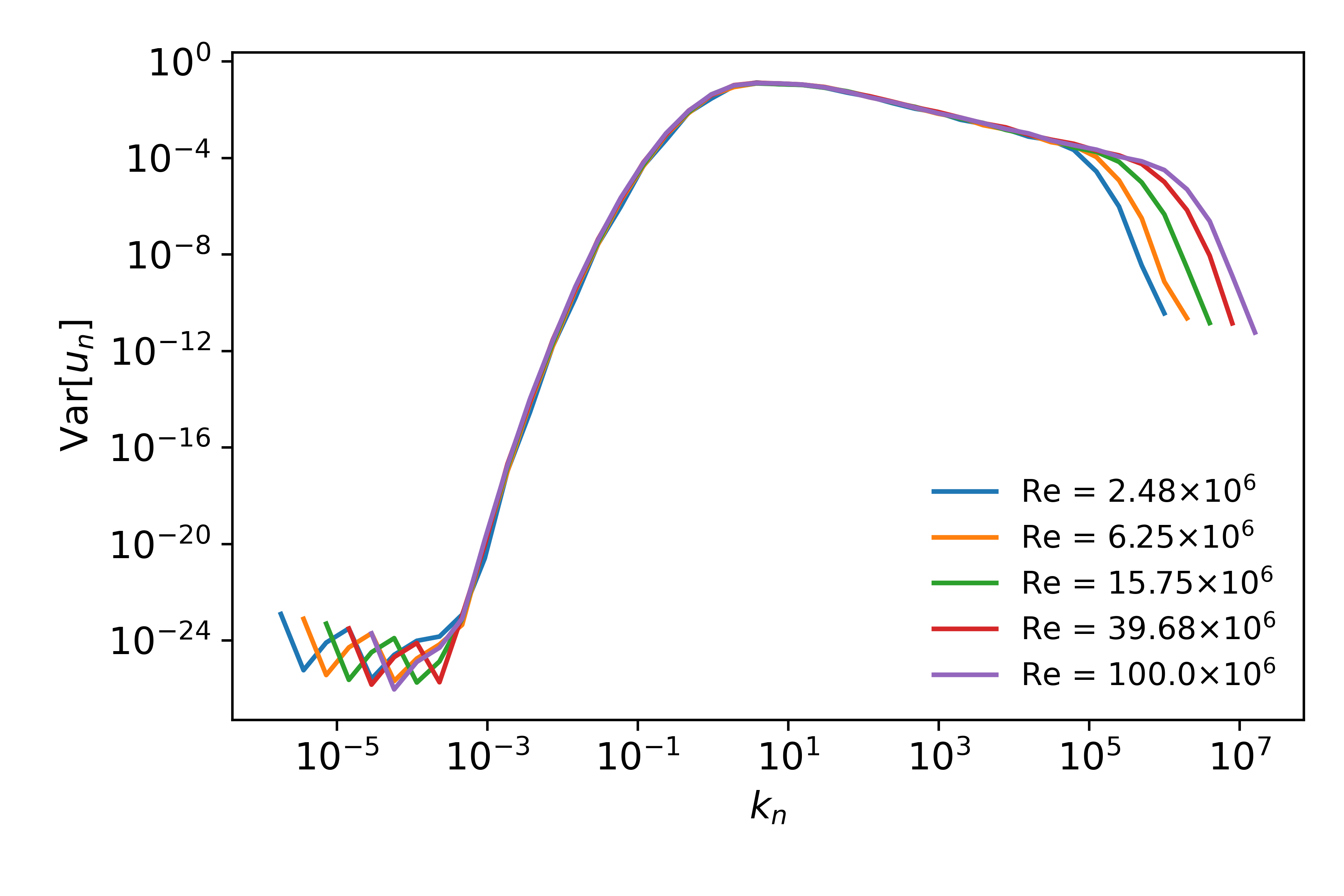}
	\caption{Variances of shell velocities averaged over the noise ensemble for the K41 initial datum, plotted as a function of wavenumber in the time-inertial units scaled with $\ell_o=\varepsilon^{1/2}t_0^{3/2}$ and $u_o=\varepsilon^{1/2}t_o^{1/2},$ for several Reynolds numbers.}
	\label{stoch-wave}
\end{figure}

Although the K41 solution does not have a natural cutoff, we had to introduce one for numerical purposes, employing 40 shells in our computations. In order for the inviscid solution to remain stationary with a cutoff, forcing can be introduced at the two shells that correspond to the largest scales, which are the only shells that are directly affected by the introduction of the cutoff. Dynamics at these two shells is given by
\be \partial_t u_0 = 0 = B_0 [u_i] +f_0 = i  k_{1} u_{2} u_{1}^* +f_0 \ee
\be \partial_t u_1 = 0 = B_1 [u_i] +f_1 = i \Big( k_{2} u_{3} u_{2}^*- 1/2 k_{1} u_{2} u_{0}^*  \Big) +f_1 \ee
Hence the forcing that ensures stationarity is the following
\be f_0=-i  k_{1} u_{2} u_{1}^* \label{force1} \ee
\be f_1 =- i \Big( k_{2} u_{3} u_{2}^*- 1/2 k_{1} u_{2} u_{0}^*  \Big) \label{force2} \ee
These values of forcing held constant were used in the whole simulation with K41 initial state.

To illustrate that, beyond the initial transient, stochastic evolution from the K41 state is self-similar, we plot variances of shell velocities across the noise ensemble as a function of wavenumber for several Reynolds numbers in time-inertial units (see Fig. \ref{stoch-wave}). Outside the dissipation range and the region close to the numerical large-scale cutoff the data points collapse to a very good approximation, which provides evidence of self-similarity of stochastic dynamics. An even more refined evidence in terms of probability distributions of the absolute values of shell velocities is given in the main text.

\section{Burst initial datum}

The ``burst'' state was generated to represent a quasi-singular state that naturally arises in the dynamics of Sabra model. Because of the very high Reynolds number required, the state was not generated via direct dynamics, but rather by a sequence of three steps. Each step corresponds to evolution of 40 shells 
with decreasing viscosity $\nu = 10^6, \ 10^3, \ 1$. At the first step the dynamics was carried out for $10$ large eddy turnover times $L/U$. The second and third steps were carried out for $10^4$ and $10^3$ viscous times respectively. This was done to ensure that the small scales are representative of dynamics with the corresponding viscosity, while keeping overall simulation time manageable.
At $\nu=1$ the simulation had only 5 shells in the dissipation range, which is marginally adequate for calculating statistical steady states but suffices to construct a single realization. 
In our subsequent simulation with thermal noise the randomness washes out the fine details of the deterministic evolution in the dissipation range and reduces the resolution requirements \cite{bandak2022dissipation}. Furthermore, the effects we are examining and measuring occur in the inertial range, which is well-resolved.

Using this procedure we identified the following ``burst'' state:
\begin{multline}
	u=[2.70338267\times 10^{3} - 4.29501775\times 10^{3} i, -5.99644021\times 10^{2} - 1.08362024\times 10^{3} i,
	-1.73526401\times 10^{2} - 8.70031976\times 10^{2} i,\\ -6.62990674\times 10^{2} + 4.43382806\times 10^{2} i,
	3.50125557\times 10^{2} + 1.94712011\times 10^{2} i, -2.90199477\times 10^{2} - 1.47957982\times 10^{2} i,\\
	1.99604442\times 10^{1} + 3.17275261\times 10^{2} i, -3.87134201\times 10^{1} - 6.38137074\times 10^{1} i,
	-6.81213766\times 10^{1} + 2.00246718\times 10^{2} i,\\ -3.05298025\times 10^{2} + 9.49052891\times 10^{1} i,
	4.02510712\times 10^{2} + 8.48877540\times 10^{1} i, -2.93971753\times 10^{2} - 1.75452647\times 10^{2} i,\\
	-1.65022863\times 10^{2} - 2.13115578\times 10^{2} i, -2.33841272\times 10^{2} + 1.26798430\times 10^{2} i,
	-6.08830647\times 10^{1} + 1.50048655\times 10^{2} i,\\ -1.28391586\times 10^{1} - 1.20865013\times 10^{2} i,
	1.98177167\times 10^{1} + 1.51201621\times 10^{2} i, -1.47897620\times 10^{1} + 1.33319974\times 10^{2} i,\\
	-2.35708969\times 10^{-1}- 1.15215796\times 10^{2} i, -6.96084174\times 10^{1} + 6.92793032\times 10^{1} i,
	-7.05189999\times 10^{1} - 5.46971275\times 10^{1} i,\\ -4.89715996\times 10^{1} + 3.74218157\times 10^{1} i,
	-3.41976233\times 10^{1} + 4.18048345\times 10^{1} i,  6.54699804\times 10^{0} + 1.35739046\times 10^{1} i,\\
	1.50962455\times 10^{1} + 2.39918367\times 10^{1} i, -2.04000328\times 10^{1} - 3.26894560\times 10^{1} i,
	1.65019396\times 10^{1} - 3.49312633\times 10^{0} i,\\ -2.94633411\times 10^{-1}- 7.39808658\times 10^{0} i,
	2.77261064\times 10^{0} + 1.71067071\times 10^{0} i,  4.92888927\times 10^{0} + 1.19176886\times 10^{0} i,\\
	-1.09264397\times 10^{0} - 2.89297482\times 10^{0} i,  1.92793947\times 10^{0} + 3.61432832\times 10^{0} i,
	-2.40679385\times 10^{0} + 2.32440175\times 10^{0} i,\\  2.66376041\times 10^{0} +1.05559511\times 10^{-1} i,
	4.77644022\times 10^{-1}- 2.06105535\times 10^{0} i,  7.31776213\times 10^{-1}+6.51759355\times 10^{-1} i,\\
	8.37075637\times 10^{-2}+1.93103368\times 10^{-1} i, -1.07909657\times 10^{-2}-4.72826940\times 10^{-3} i,
	7.46353392\times 10^{-5}+3.30852429\times 10^{-6} i, \\ 6.18596728\times 10^{-9}-1.20561074\times 10^{-8} i].
	\nonumber
\end{multline}

\begin{figure}
	\centering
	\includegraphics[width=0.45\linewidth]{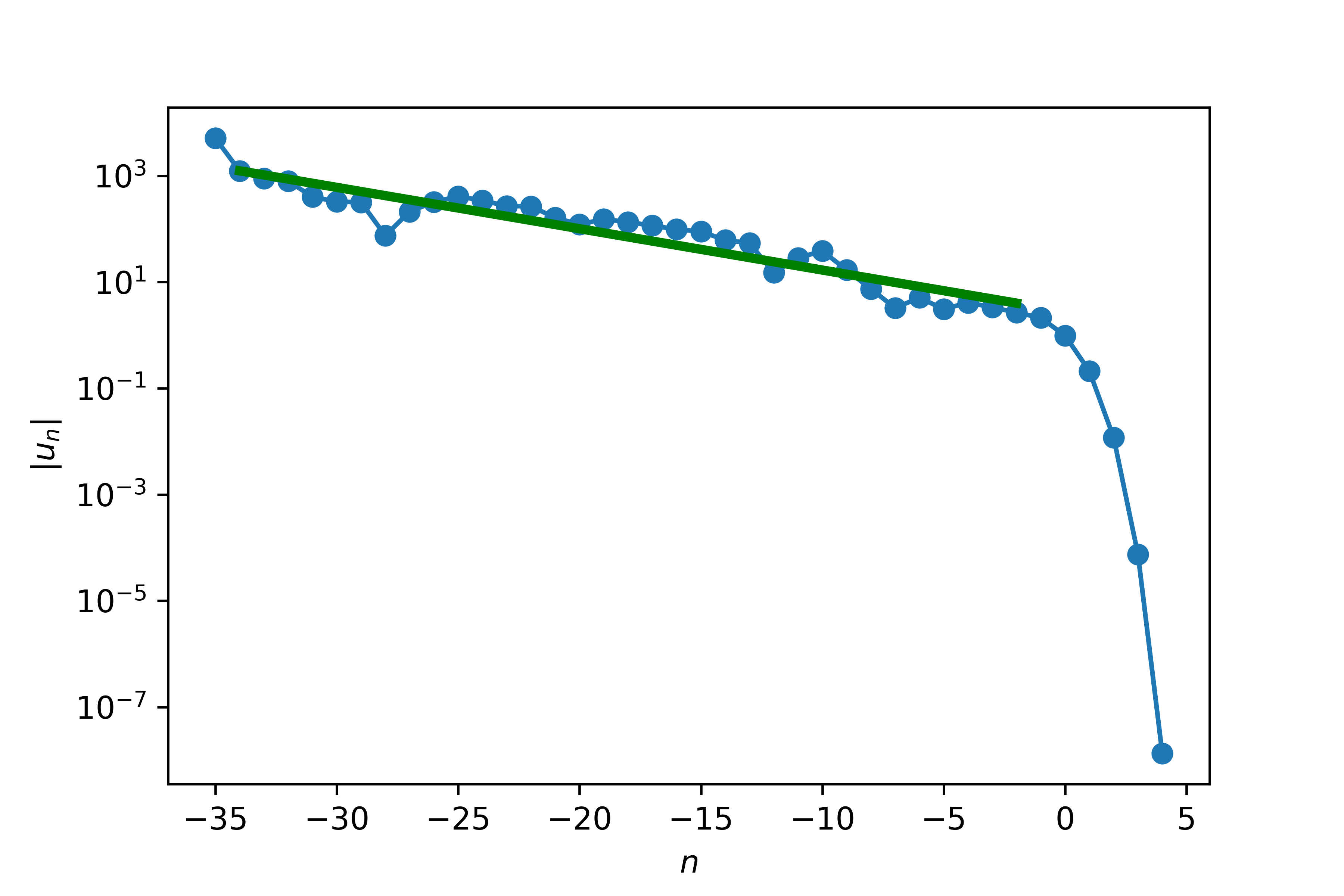}
	\caption{Absolute values of shell velocities of a burst initial state as a function of shell index $n = \mathrm{log}_2 k_n + \mathrm{log}_2 L$, and a power-law fit with exponent $-0.258$.}
	\label{burst-holder}
\end{figure}

In Fig. \ref{burst-holder} we depict the absolute values of the burst state as a function of shell index. The scaling of the absolute values is roughly described by the H\"{o}lder exponent 
$h \simeq 1/4$. The values of $u_0, u_1, u_2$ given above were used to calculate the constant forcing in the simulation of the dynamics of the burst state from \eqref{force1} and \eqref{force2}.

\section{Supplement to Figure 1(c):  Varying Reynolds number for the ``burst'' initial condition.}
We explain here briefly how we performed the numerical simulations for the ``burst'' initial datum at different Reynolds numbers $Re=UL/\nu$ obtained by varying $\nu.$ The main computational issue in performing these integrations was the choice of time-step $\Delta t.$ The scaling of the viscous wavenumber is given by
\be k_{\nu} \propto (\epsilon \nu^{-3} )^{1/4}. \ee
Therefore increasing of viscosity by a factor of $2^5$ decreases the viscous wavenumber by a factor of $2^{15/4}$, which is equivalent to a shift of index $n$ by $-3.75$. Given that our slaved integration scheme requires $\nu k_n^2 \Delta t$ to be small, decreasing our truncation index by $4$ allowed us to increase the timestep in our simulation as follows
\be \Delta t' \propto \frac{1}{(2 ^5 \nu ) k_{N-4}^2} = 2^3 \Delta t. \ee
The nominal Reynolds numbers achieved by this approach are extremely large, up to $10^{14}.$ However, for the study of spontaneous statistics at shell $n=18$ in Figure 1, panel (c) these large Reynolds numbers are unrealistically inflated, because the modes with shell numbers $n<18$ are essentially frozen on the time-scales of interest and play no role. We have therefore defined Reynolds number for the purpose of Figure 1(c) by the length-scale $\ell_n$ and root-mean-square velocity $U_n$ that correspond to the shell of interest, that is $n=18$. The final-time r.m.s. velocity is calculated by an average over noise realizations.

\section{Supplement to Figure 1: Control experiment without thermal noise}

\begin{figure}
	\centering
	\includegraphics[width=0.5\linewidth]{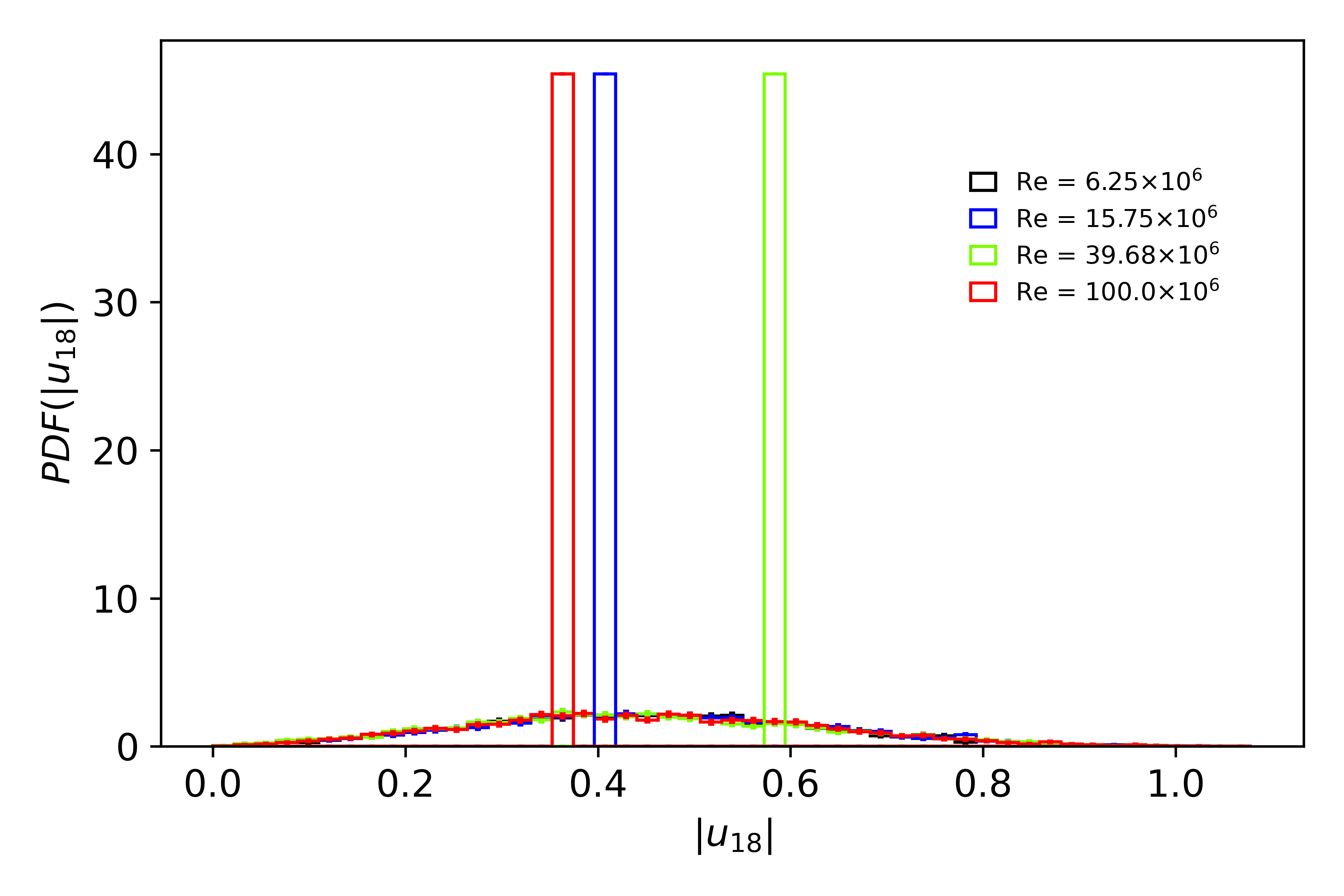}
	\caption{Transition probability density function for the absolute values $|u_n|$ at a single fixed shellnumber $n = 18$ and time $t_f = 1$,
		exactly as in Fig.1(a) in the main text, but with thermal noise turned off. 
		%There is exactly one realization for each of the four Reynolds numbers, so that the PDF's are delta functions represented by a single occupied bin. 
		The PDF's for $Re=6.25\times 10^6$ and $Re=10^8$ are 
		identical and lie on top of each other. For comparison, the original PDF's from Fig.~1(a) in the main text are replotted here as well.}
	\label{control}
\end{figure}

\textcolor{black}{
	As a simple control experiment, we have recalculated the transition PDF’s plotted in Fig.~1(a), but now with thermal noise set identically to zero.
	The results are plotted in Fig.~\ref{control}, where it can be seen that the PDF's become Dirac delta functions. This is expected because, in the 
	absence of external noise, we are solving deterministic ODE’s in finitely many variables with deterministic initial data. There is thus exactly one 
	solution of the initial-value problem for each Reynolds number and the corresponding unique value of $|u_{18}|$ lies in a single bin of the histogram 
	plotted below for the control experiment. 
	There is no discernible limit as $Re\to\infty.$ In fact, we observe periodicity in $Re,$ explaining why the result for $Re=6.25\times 10^6$ and $Re=10^8$ coincide. 
	Our observations are consistent with rigorous results of \cite{constantin2007regularity,mailybaev2016spontaneous}, who proved that solutions of the viscous Sabra shell model will in general 
	converge as $Re\to\infty$ to a weak solution of the inviscid Sabra model with infinitely many shells. However, such convergence is only obtained along a 
	suitable subsequence of Reynolds numbers, and the limiting solution may depend on the subsequence. Fig.~\ref{control} illustrates how essential 
	is the thermal noise (or some external noise) for the results in Fig.~1 of the main text.}

\textcolor{black}{
	Although the randomness appearing in Fig.~1 of the main text is due to the thermal noise, it is not a direct effect of that noise. As discussed in 
	the main text of the letter, the observed randomness is due instead to the “inverse error cascade” of Lorenz, which propagates the effect of thermal 
	noise from the dissipation range up to any inertial-range scale in a few local turnover times. To explain why the direct effects of thermal 
	noise are negligible, let us consider the parameter values $\alpha=0,$ $n=18$, and $t_f=1$ used in Fig.~1(a). 
	Note that the direct thermal noise effects are even smaller for the other cases illustrated in Fig.~1(b,c). The r.m.s. velocity due to the direct effect 
	of thermal noise is $u^\theta_{rms}=(2\Theta k_n^{2+\alpha}t_f)^{1/2}$ with $\Theta=Re^{-\beta}\theta_\eta$, where for all cases considered we    
	took the value $\theta_\eta=2.83\times 10^{-8}$ typical of the atmospheric boundary layer. Thus, for $n=18$ at the lowest considered Reynolds number, 
	$Re=6.25\times 10^6,$ we have $u^\theta_{rms}=4.99\times 10^{-4}$ and at the highest Reynolds number, $Re=10^8,$ we have $u^\theta_{rms}=
	6.24\times 10^{-5}$. These values are obviously negligible compared with the r.m.s velocity of the spontaneous statistics exhibited in Fig.1(a). 
	These large spontaneous values originate  from the stochastic wave shown in Fig.~2 of the main text, which arrives to shellnumber $n=18$ at a dimensionless time 
	$t=3.4(2^{18})^{-2/3}=8.3\times 10^{-4}$, asymptotically independent of $Re$. The large spontaneous fluctuations seen in Fig.~1(a) are thus 
	due to thermal noise effects in the dissipation range propagated up into the inertial range by nonlinear error cascade, and not due 
	to direct local effects of thermal noise.}   

\section{Supplement to Figure 1: Energy flux probability density functions}
\begin{figure}
	\centering
	\begin{subfigure}{0.45\linewidth}
		\includegraphics[width=\linewidth]{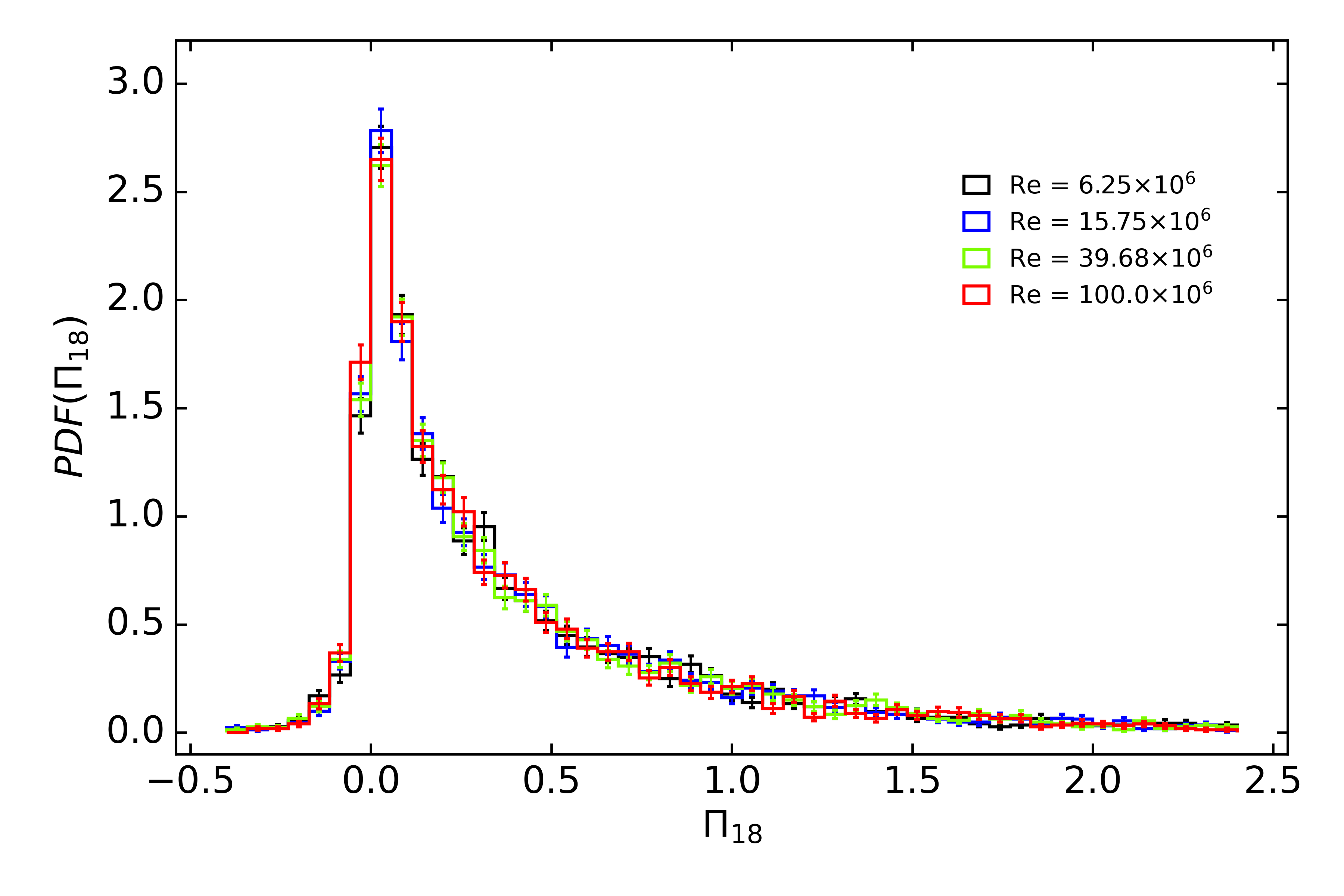}
		\caption{Self-similar initial state with noise scaling $\alpha = 0.$}
	\end{subfigure}
	\begin{subfigure}{0.45\linewidth}
		\includegraphics[width=\linewidth]{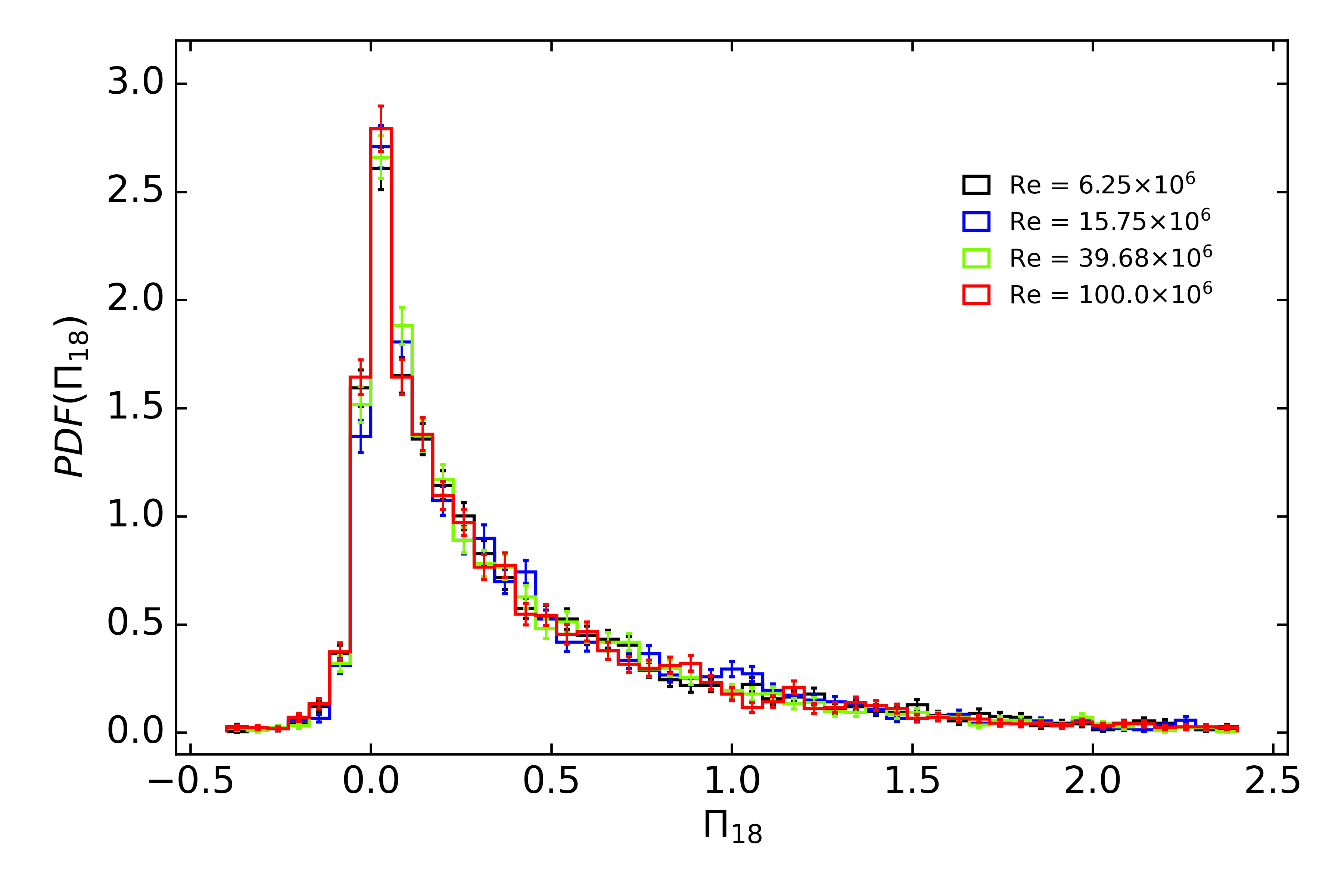}
		\caption{Self-similar initial state with noise scaling $\alpha = 3.$}
	\end{subfigure}
	\caption{Probability density functions of the energy flux $\Pi_{18}$ for a fixed wavenumber and time in inertial units for several Reynolds numbers. All the errors are estimated as standard errors using the bootstrap method.}
	\label{pdf-flux}
\end{figure}

In Fig.1 of the main text we plotted probability density functions of $|u_n|$ for multiple Reynolds numbers as evidence of spontaneous stochasticity. To provide more evidence,
Fig. \ref{pdf-flux} plots probability density functions of energy flux $\Pi_n = \frac{1}{2} T_{n-1} + T_{n}$ with $T_n =  k_n \mathrm{Im}\{ u_{n}^* u_{n+1}^* u_{n+2} \}.$
These appear as well to converge to limiting distributions with increasing $Re.$

\section{Dimensional analysis of the self-similar stochastic wave}

\begin{figure}
	\centering
	\begin{subfigure}{0.45\linewidth}
		\includegraphics[width=\linewidth]{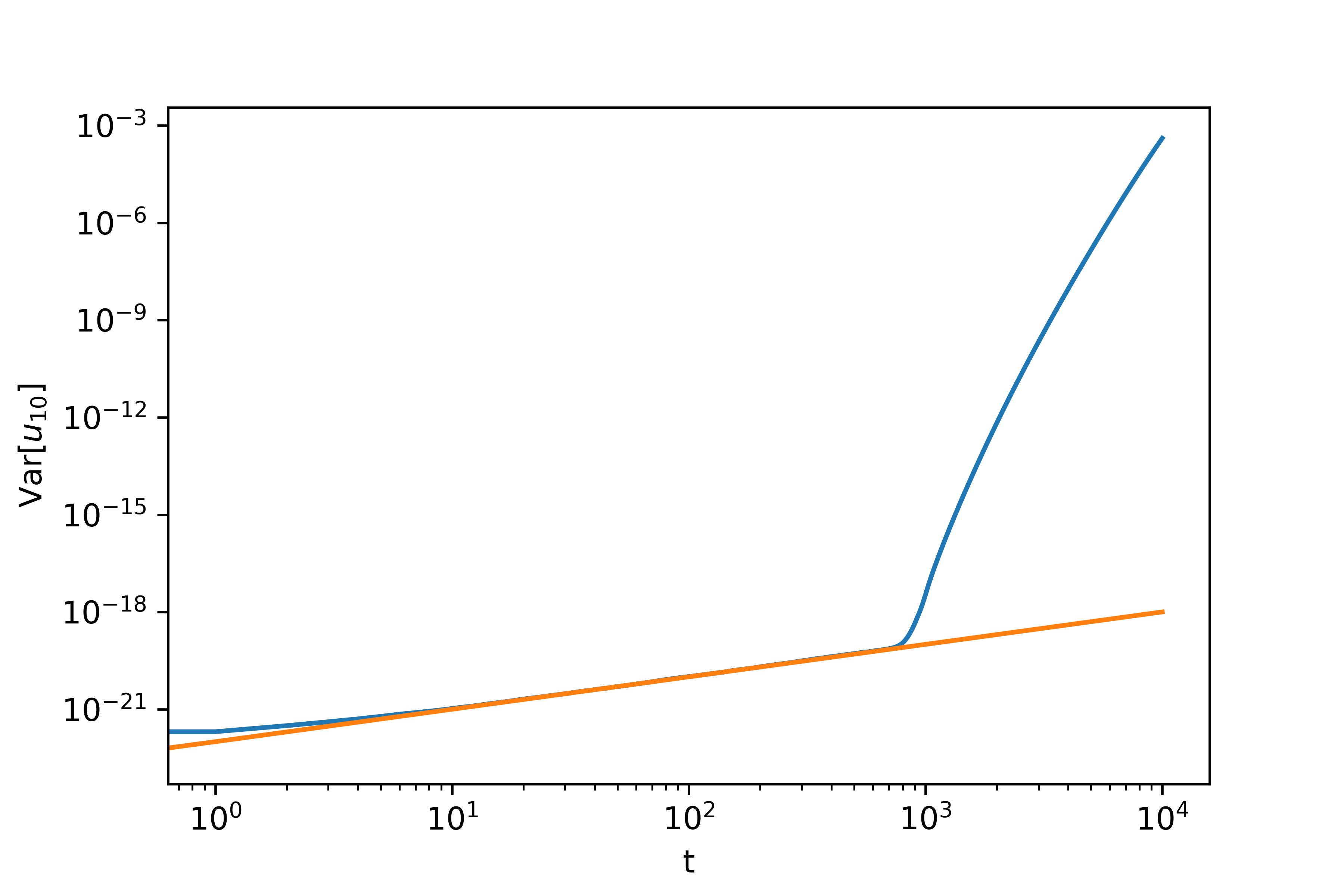}
		\caption{$n = 10$}
	\end{subfigure}
	\begin{subfigure}{0.45\linewidth}
		\includegraphics[width=\linewidth]{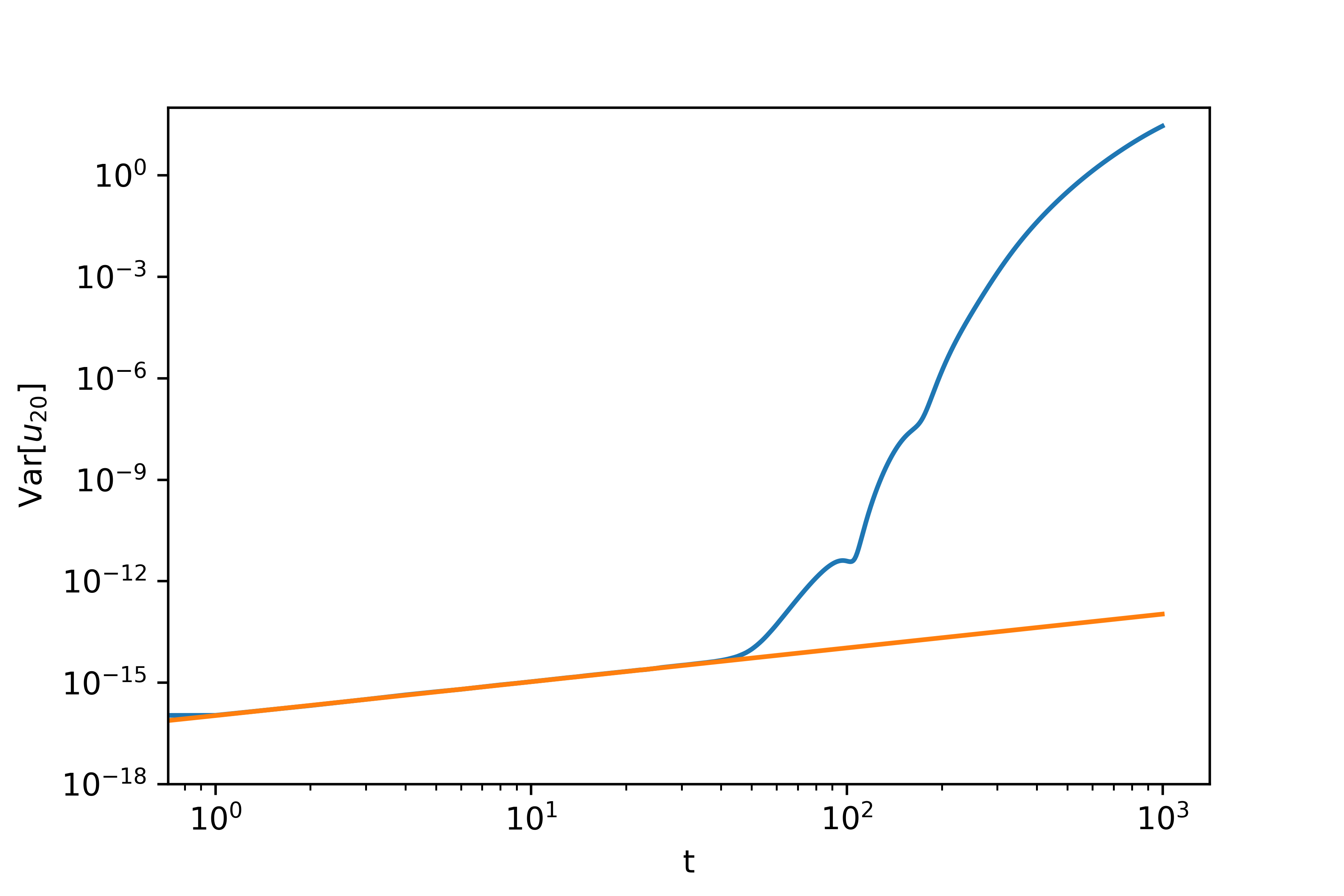}
		\caption{$n = 20$}
	\end{subfigure}
	\caption{Growth of variance $\mathrm{Var}[u_n]$ (blue line) as a function of time. For comparison we also plot the function $4 \theta_\eta k_n^2 t$ (orange line), which clearly coincides with $\mathrm{Var}[u_n]$ for early times once the variance supersedes the roundoff error.} \label{initial-linear}
\end{figure}

\begin{figure*}
	\centering
	\begin{subfigure}{0.4\linewidth}
		\includegraphics[width=\linewidth]{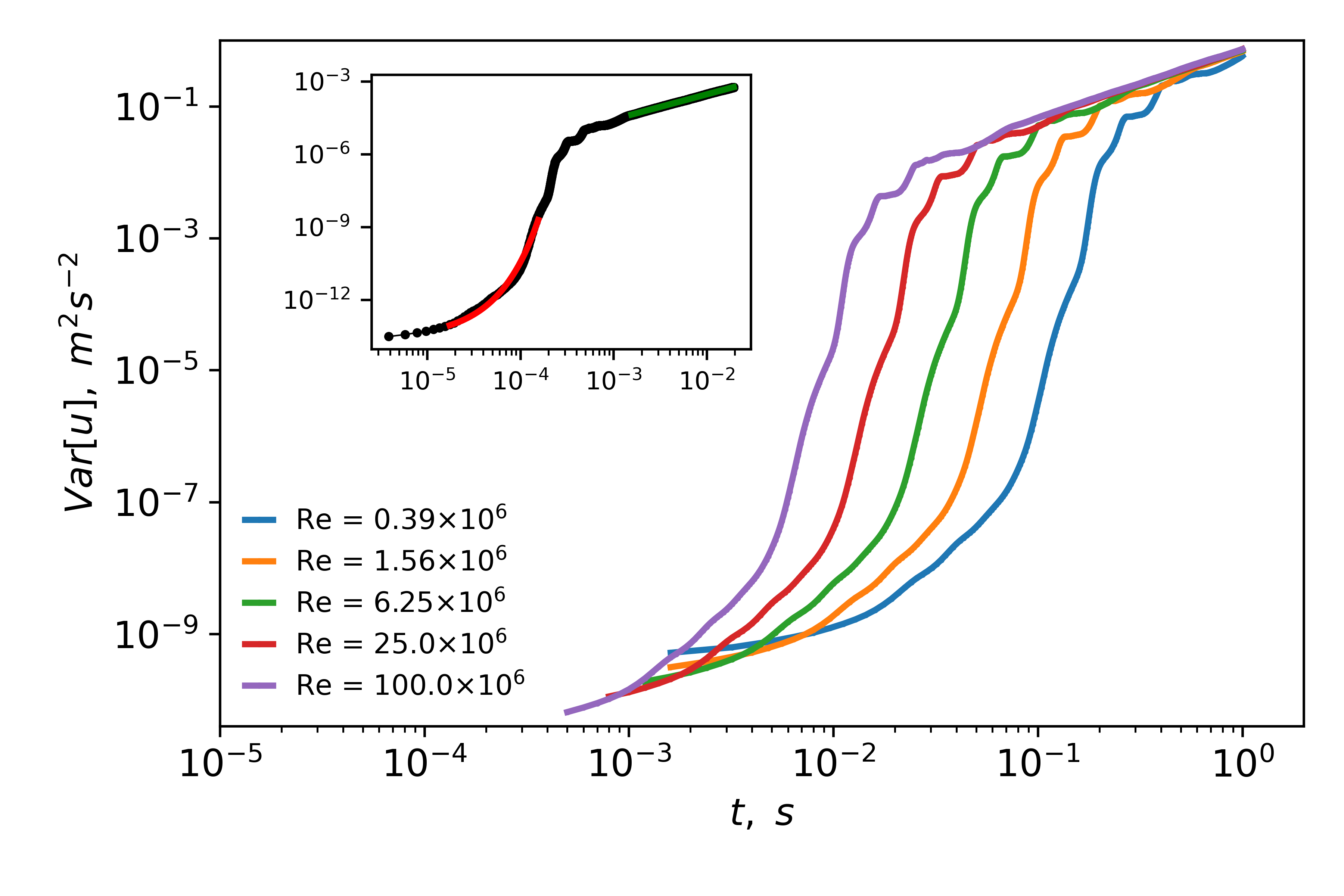}
		\caption{Stationary K41 initial state.}
	\end{subfigure}
	\begin{subfigure}{0.4\linewidth}
		\includegraphics[width=\linewidth]{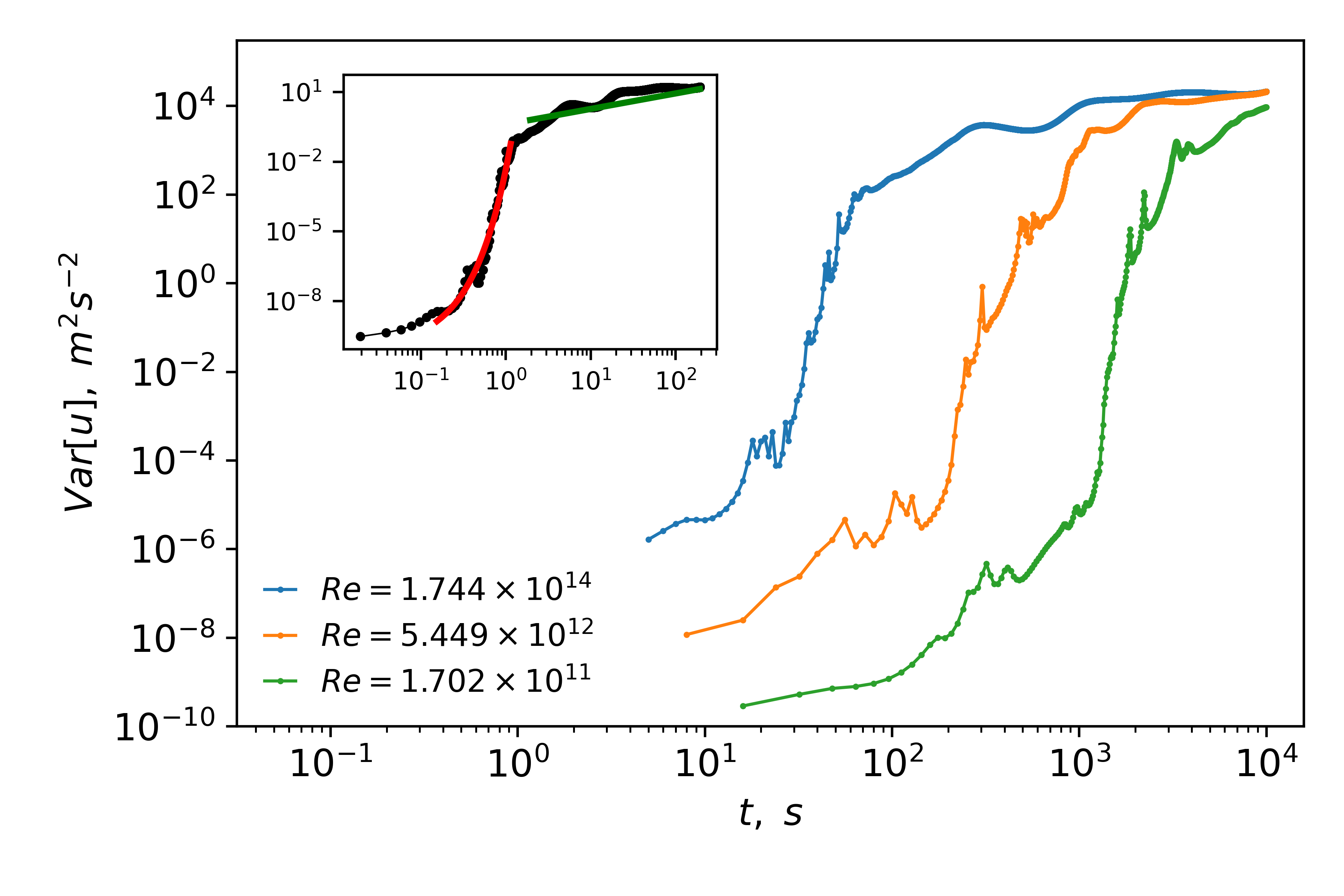}
		\caption{``Burst'' initial state.}
	\end{subfigure}
	\caption{Total variance growth as a function of time in inertial range units for the two initial states. The inset plots show the total variance growth as a function of time in SI units for the largest Reynolds number. The green line on the inset plots shows predicted power-law growth in the self-similar regime, and the red line on the inset plots shows exponential fit in the chaotic regime.} \label{var-growth}
\end{figure*}

\textcolor{black}{The self-similarity of the stochastic wave propagation is best visible in the evolution of the total variance $\mathrm{Var}(\bu)=\sum_n \mathrm{Var}(u_n)$ summed over all shells, with
	\be \mathrm{Var}(u_n):= \langle |u_n-\langle u_n\rangle|^2\rangle.  \lb{varn} \ee
	Here the average is over the ensemble of noise realizations. This quantity is plotted in Fig.~\ref{var-growth} for both the K41 (panel a)  and the ``burst'' (panel b) initial conditions, at several values of $Re$. Note that the thermal noise in \eqref{stoch-sabra} by itself would produce diffusive variance growth, linear in time for each shell. This linear growth is given by $\mathrm{Var}(u_n)\sim (2\theta_\eta/Re^{3/2})k_n^2 t$, or $\mathrm{Var}(u_n)\sim (4\nu k_BT/\varrho)k_n^2 t$ before non-dimensionalization, which is what we observe for each shell separately at early time. This can be seen in Fig. \ref{initial-linear} for $n=10$ and $n=20,$ but similar behavior is found for all $n.$  Note the slight deviation from linear growth near $t=0$ due to round-off error of the double precision arithmetic.}

\textcolor{black}{Exponential growth, indicated  by the red curve in the insets in Fig. \ref{var-growth}, commences when the nonlinearity starts to dominate, which is the chaotic phase predicted by Ruelle \cite{ruelle1979microscopic}. Finally, however, power-law growth ensues, to a very good approximation for the K41 initial datum and to rough approximation for the ``burst'' datum. For any exactly self-similar initial condition scaling as $u_{n} = A k_n^{-h},$ the variance could be expected to grow as a power law $\mathrm{Var}(\bu) \propto A^{\frac{2}{1-h}}t ^{\frac{2h}{1-h}}$ on dimensional grounds. This is observed in Fig. \ref{var-growth} for the K41 solution, where $h= 1/3$ gives $\mathrm{Var}(\bu) \propto \varepsilon t,$ in good agreement with the numerical results. Notice that this is the same power-law observed in previous studies of predictability of turbulence, both by analytical closure \cite{leith1972predictability} and by direct numerical simulation \cite{boffetta2001predictability,boffetta2017chaos}. For the ``burst'' solution the power-law $\mathrm{Var}(\bu) \propto t^{2/3}$ predicted for $h\doteq 1/4$ is only crudely consistent with the data. In the spirit of the multifractal model \cite{frisch1985singularity}, we conjecture that power-law growth would be recovered if the variance were averaged over many such ``burst'' initial data all selected from the statistical steady-state with the same H\"older exponent $h.$ Averaging further over a distribution of $h$-values should reproduce the anomalous growth $\langle |\delta\bu(t)|^p\rangle\sim t^{\xi_p},$ with $\xi_p$ nonlinear in $p,$ previously observed for growth of differences $\delta\bu(t)=\bu'(t)-\bu(t)$ between solutions in shell models \cite{aurell1996predictability}. }

\section{Statistics of Kolmogorov multipliers and super-universality}

\begin{figure*}
	\centering
	\begin{subfigure}{.3\linewidth}
		\includegraphics[width=\linewidth]{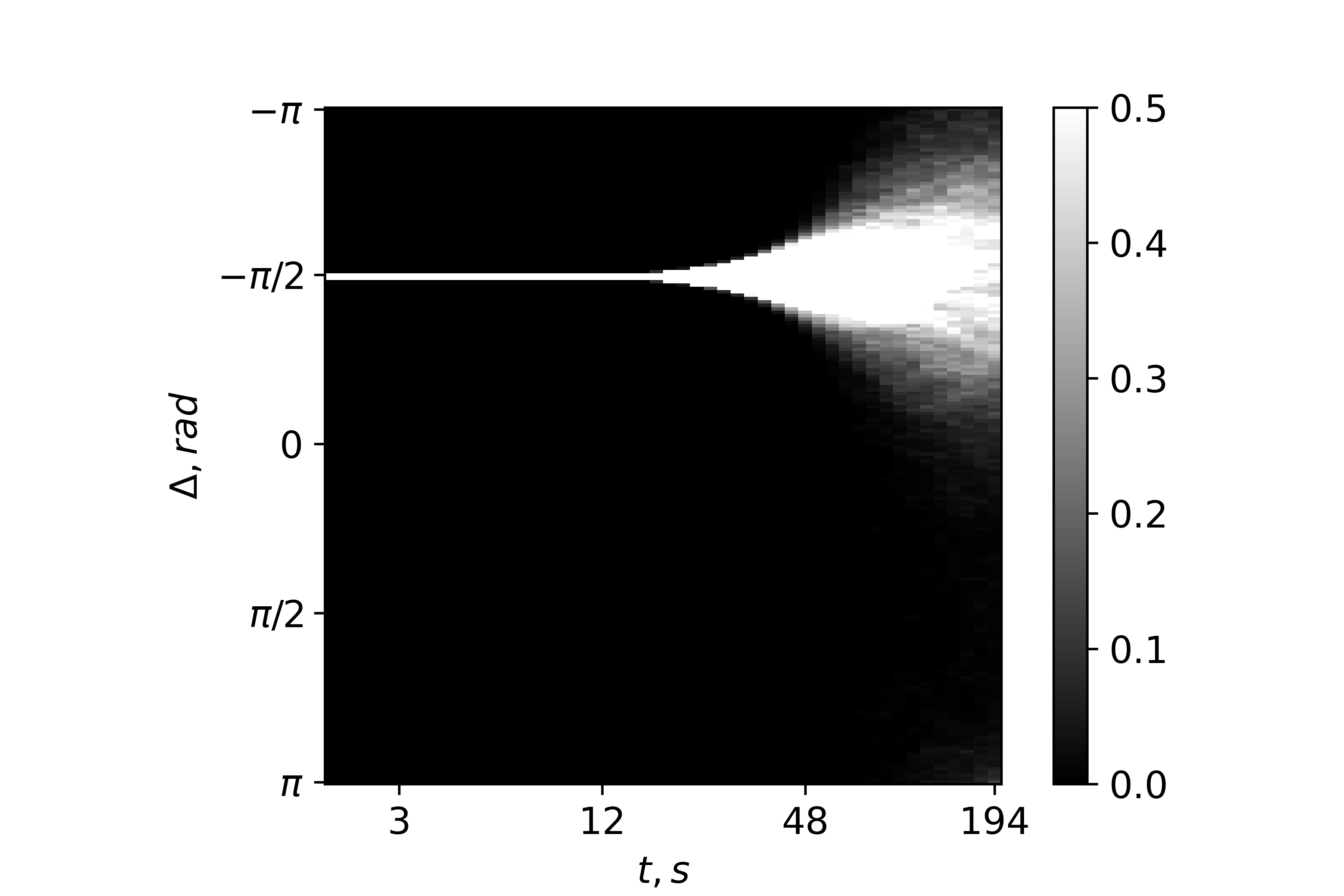}
		\caption{$n = 16$}
	\end{subfigure}
	\begin{subfigure}{.3\linewidth}
		\includegraphics[width=\linewidth]{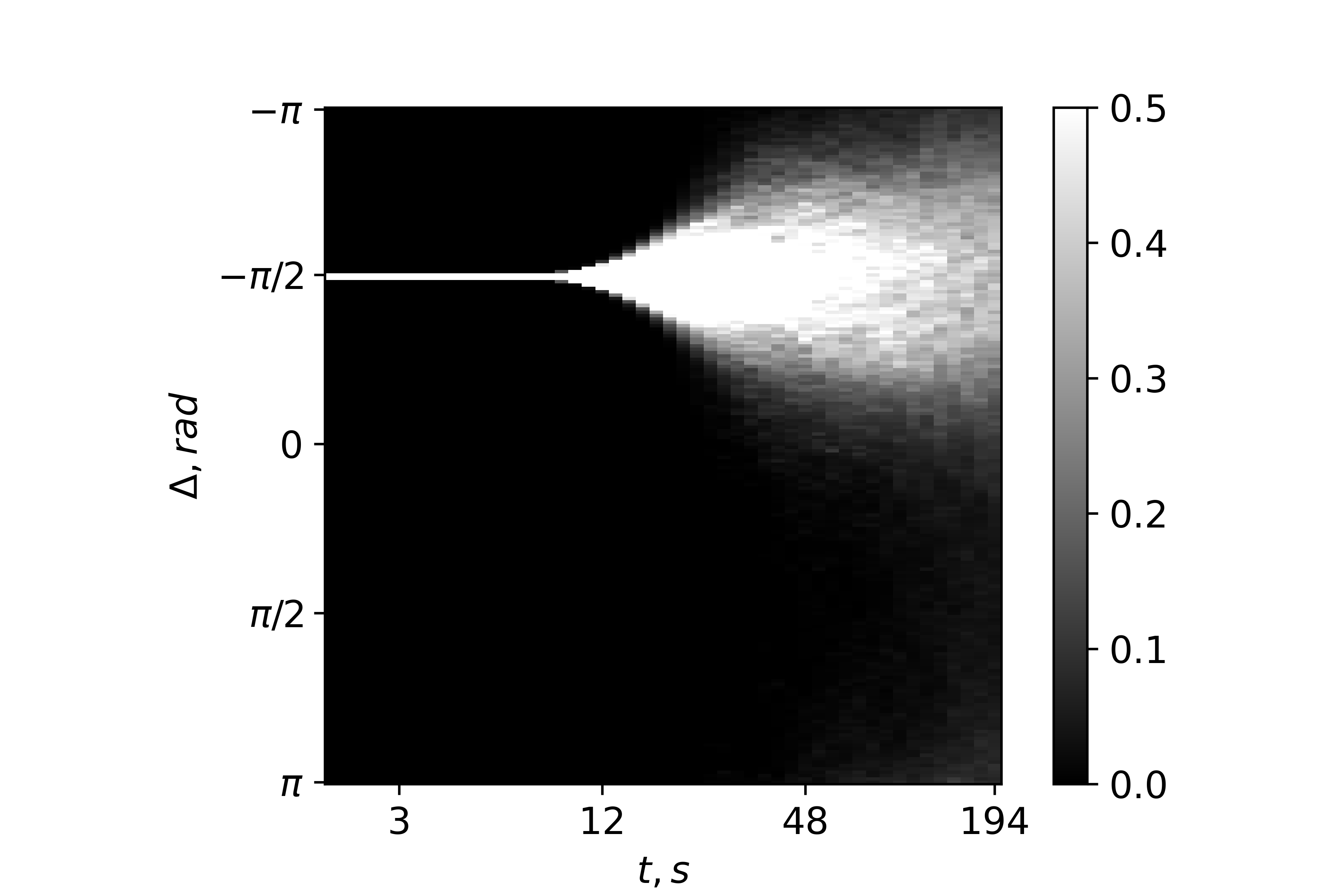}
		\caption{$n = 18$}
	\end{subfigure}
	\begin{subfigure}{.3\linewidth}
		\includegraphics[width=\linewidth]{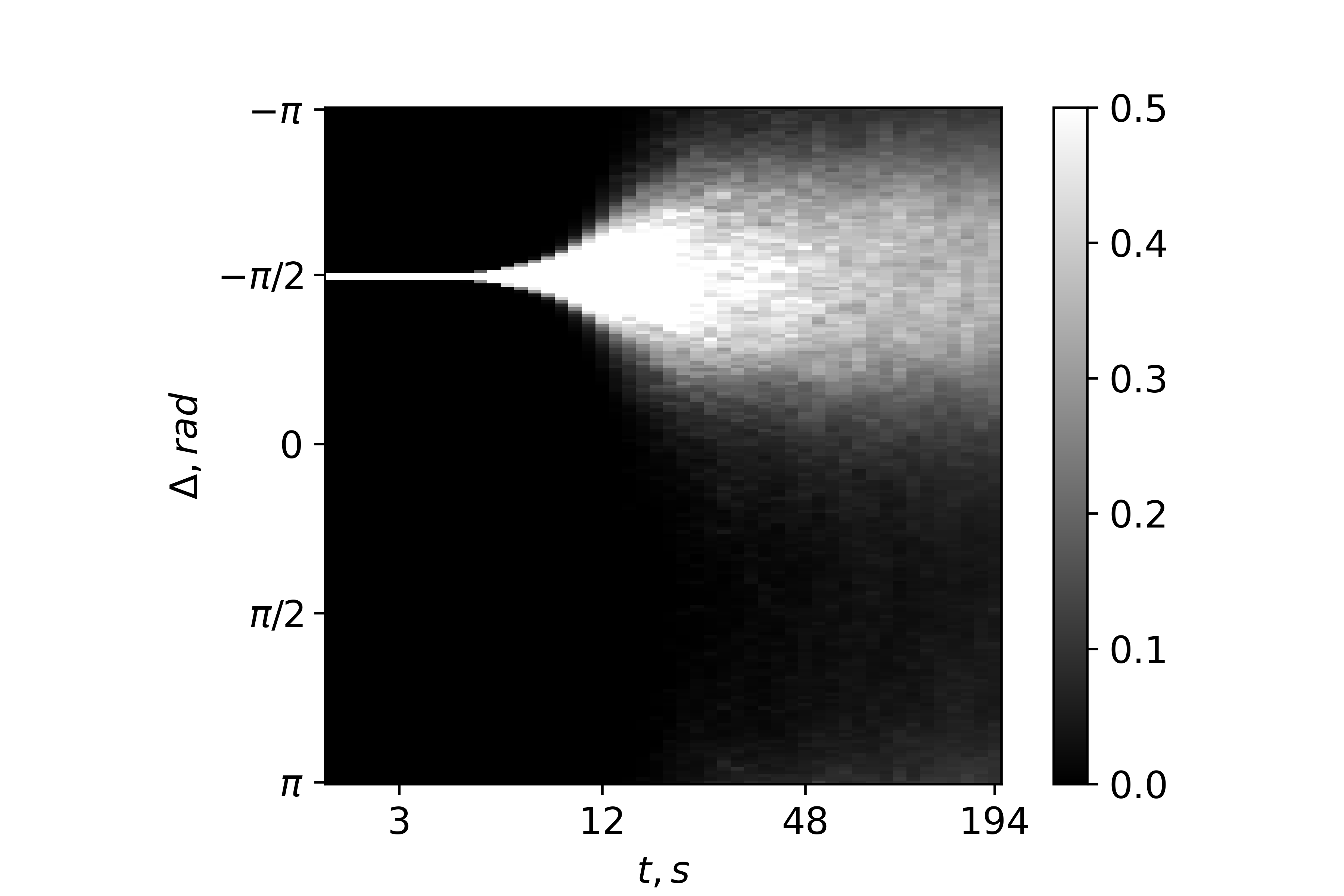}
		\caption{$n = 20$}
	\end{subfigure}
	\caption{Probability density functions of angle multipliers as a function of time in SI units for several shell indices in the inertial range. The three subfigures illustrate the self-similarity of the stochastic wave, which follows from identical probability density functions for different shells up to a time shift.} \label{pdf-mult}
\end{figure*}

\begin{figure}
	\centering
	\begin{subfigure}{0.45\linewidth}
		\includegraphics[width=\linewidth]{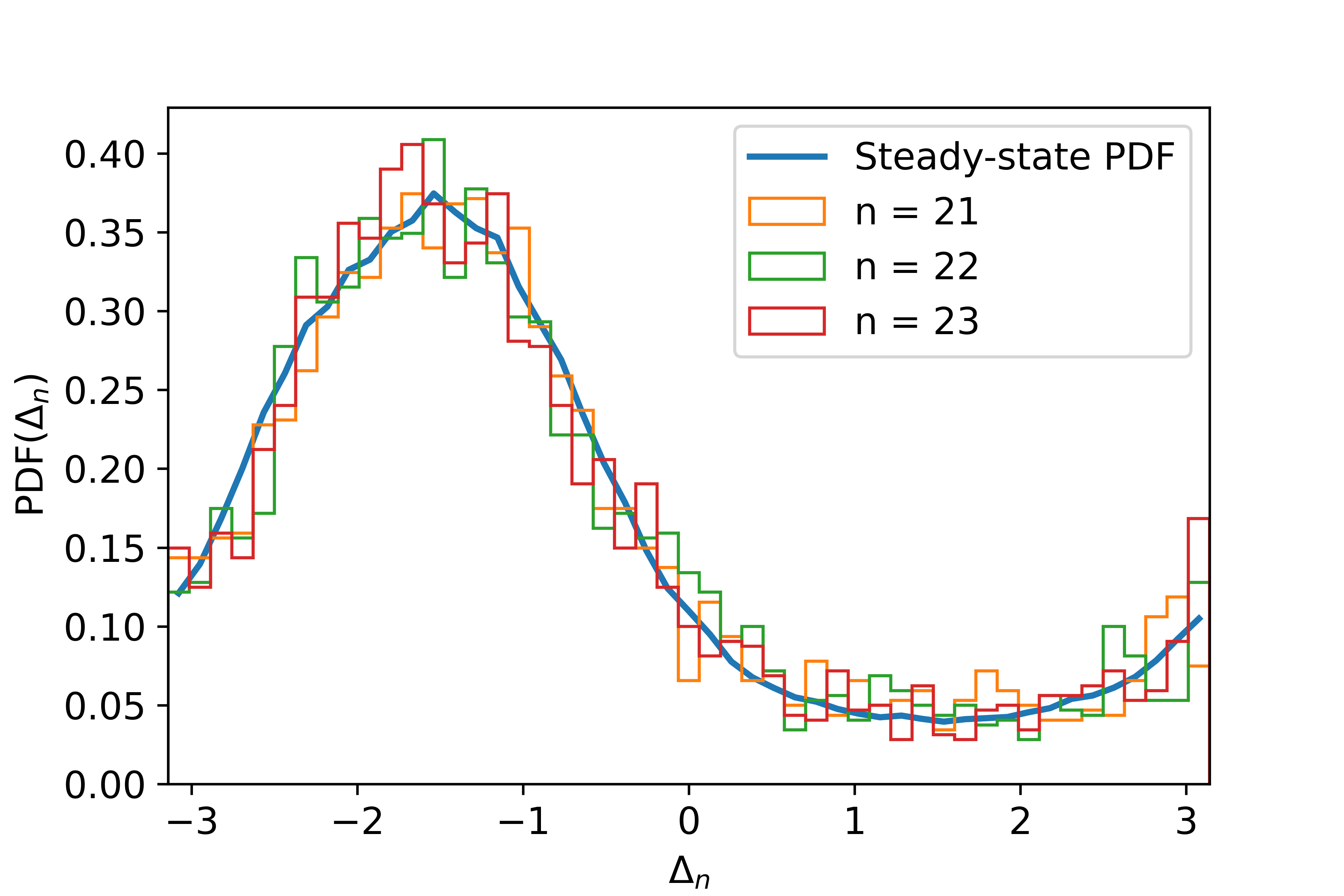}
		\caption{Angular multipliers}
	\end{subfigure}
	\begin{subfigure}{0.45\linewidth}
		\includegraphics[width=\linewidth]{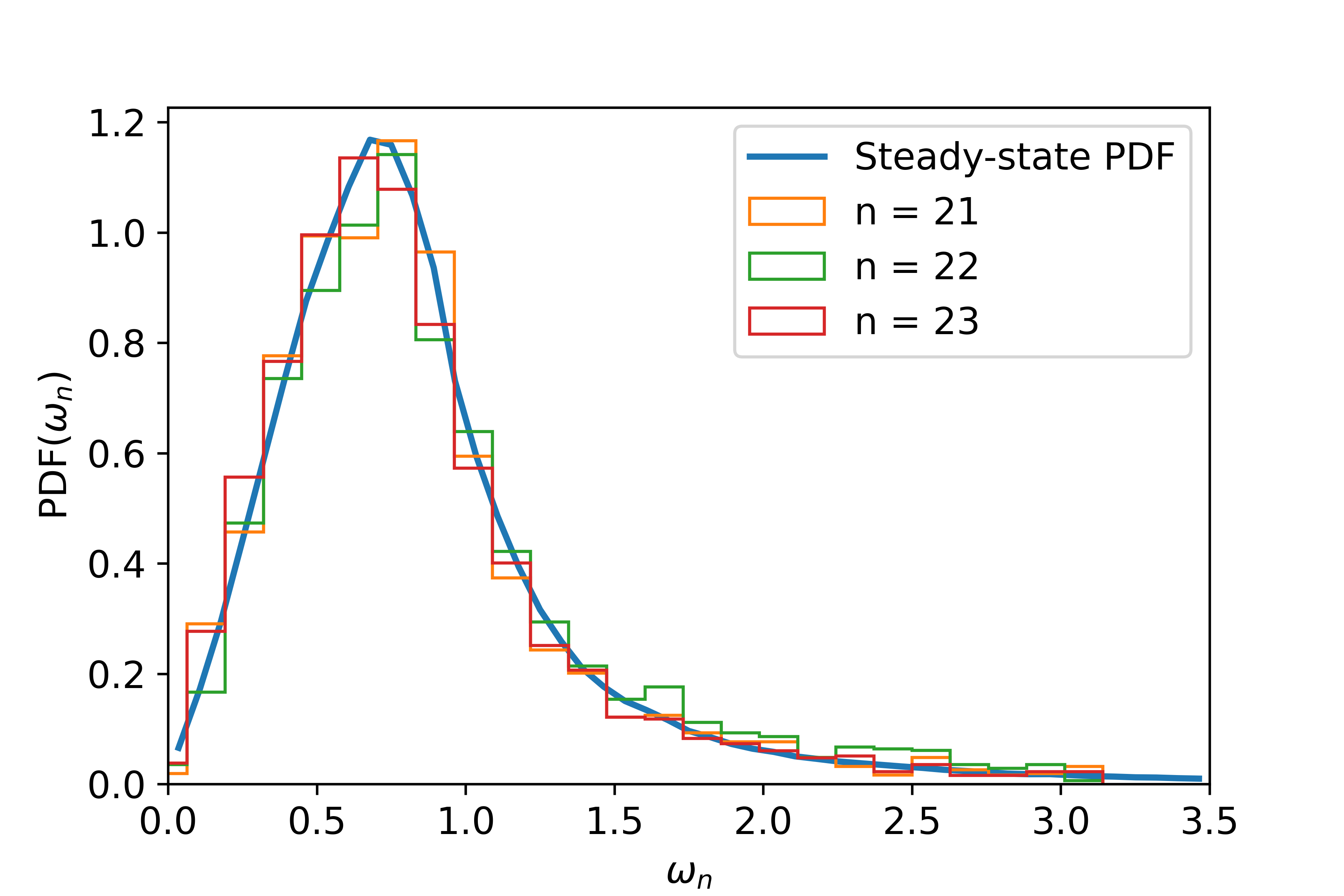}
		\caption{Magnitude multipliers}
	\end{subfigure}
	\caption{Comparison of the probability distributions of angle and magnitude multipliers in the steady-state with the probability distribution in the stochastic wake past the front.}  \label{super-univ}
\end{figure}

\textcolor{black}{Once the stochastic wave has passed a given shell $n,$ further changes at that scale are due only to the intermittent nature of the turbulent dynamics behind the front, which we find to be similar to that in the infinite-time statistical steady-state. To expose this deep similarity, we consider the  Kolmogorov multipliers \cite{kolmogorov1962refinement} for the shell amplitudes $\omega_n = |u_n|/|u_{n-1}|$ and the phase multipliers \cite{eyink2003gibbsian} $\Delta_n = \mathrm{arg} \big( u_{n-2} u_{n-1} u_n^* \big)$, which have been used previously to characterize the build-up of intermittency in the turbulent cascade. The PDF's of angle multipliers are shown in Fig. \ref{pdf-mult} for three fixed shells, $n=16,18,20,$ as a function of time. They are invariant under the shift $n \mapsto  n-1$, $t=\tau_n\mapsto t=\tau_{n-1}$, which is a consequence of the self-similar growth. Furthermore, unlike the PDFs of modal energy and energy flux, which keep evolving in time for each shell and only coincide in time-inertial units, the PDFs of multipliers saturate and become time-invariant past the front. This is indicative of the scale-invariance of the multiplier statistics, a well-known feature in the turbulent steady-state which has been explained there by a ``hidden symmetry'' \cite{mailybaev2020hidden,mailybaev2022hidden}. In fact, the multiplier PDFs in the stochastic wake of the traveling wave are indistinguishable from those in the infinite-time steady state. This is what we call {\it super-universality} of the multiplier statistics, distinct from the usual universality of the spontaneous statistics, that is, its independence from the precise regularization and the small-scale noise that triggers it \cite{mailybaev2016spontaneously}. As evidence of super-universality in the Sabra model, we compare the probability distribution functions of magnitude and angle multipliers in the wake of the stochastic front with those same probability densities observed in a turbulent steady-state from our previous study \cite{bandak2022dissipation}. As seen in Fig. \ref{super-univ}, there is excellent agreement. As a consequence, we expect that structure functions $\langle |u_n|^p\rangle$ calculated in the stochastic wake will exhibit anomalous scaling $\sim (\varepsilon/k_n)^{p/3} (k_n\ell(t))^{-\delta\zeta_p}$, exactly as in the steady-state cascade but with usual outer length $L$ replaced by $\ell(t).$ Super-universality thus implies that characteristic turbulent statistics are reached in finite time, unlike the universal steady-state distributions in low-dimensional chaotic systems which require an infinite time limit.}

\section{Supplement to Figure 2: Lorenz plot for ``burst'' initial datum}
\begin{figure}
	\centering
	\includegraphics[width=0.45\linewidth]{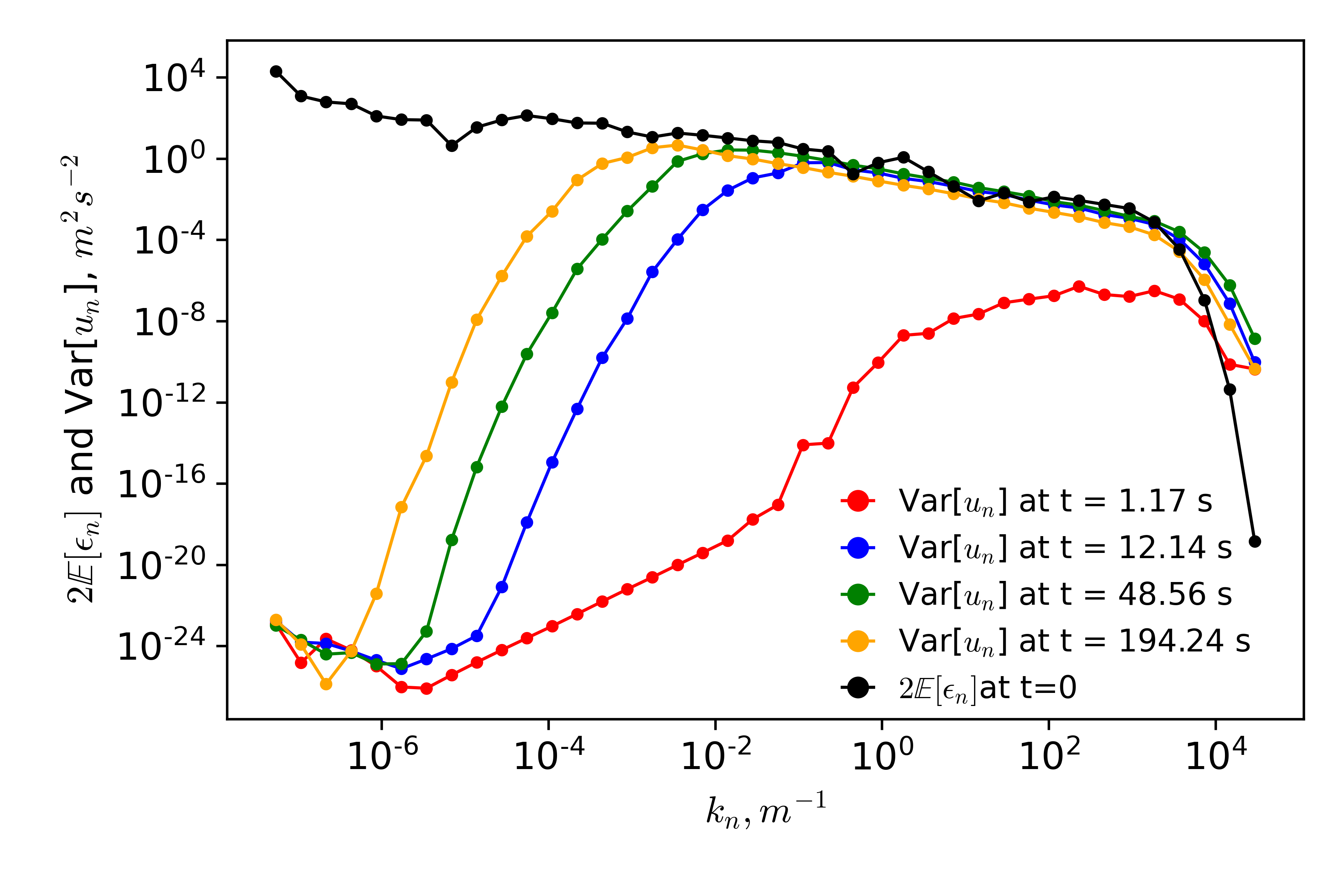}
	\caption{{Twice ensemble average energy $\mathbb{E}[\epsilon_n]$ (\textcolor{black}{$\bullet$}) for $\epsilon_n=\frac{1}{2}|u_n|^2$ and velocity variances (\textcolor{red}{$\bullet$},\textcolor{blue}{$\bullet$},\textcolor{green}{$\bullet$},\textcolor{orange}{$\bullet$}) across the ensemble as a function of wavenumber in SI units for 4 increasing times. The smallest time in the variances plots the initial transient and the subsequent three times show the propagation of the stochastic front across the inertial range towards large scales. $2\mathbb{E}[\epsilon_n]$ of the initial state is provided for comparison with the Fig. 2 of the main text.}} \label{lorenz-wild}
\end{figure}

In Fig.2 of the main text we presented a plot analogous to that of Lorenz (1969) \cite{lorenz1969predictability}, in his Figure 2, for our shell model with K41 initial data.  Fig. \ref{lorenz-wild} shows the similar plot for the ``burst'' initial datum, exhibiting a similar spread of stochasticity across scales.

\section{Theoretical Estimate of the Condition for Eulerian Spontaneous Stochasticity.}
\textcolor{black}{Building on the earlier work of Ruelle \cite{ruelle1979microscopic} and Lorenz  \cite{lorenz1969predictability}, we can provide a theoretical estimate of the randomization times $t_r(n)$, which will also allow us to estimate how we can take the double limit $\Theta, \mathrm{Re}^{-1} \rightarrow 0$ to arrive at the spontaneously stochastic state. According to Ruelle's estimate \cite{ruelle1979microscopic} the time it takes for a thermally-triggered disturbance $\Delta$ of velocity to become significant at the Kolmogorov scale $\eta$ of a turbulent flow is $t_R \sim (\nu/\varepsilon)^{1/2} \mathrm{log}(\theta_\eta^{-1/2} )$. The growth of a disturbance in Ruelle's regime is exponential with Lyapunov exponent that corresponds to the Kolmogorov scale, but after it reaches the inertial range it enters the Lorenz regime, and grows in a self-similar way across the cascade. If we assume complete self-similarity, then by dimensional argument the disturbance grows as $\Delta(t) \sim \varepsilon^{1/2} (t-t_o)^{1/2}$, and its length-scale propagated up the cascade is $\ell(t)\sim \varepsilon^{1/2}(t-t_o)^{3/2}$. Both of these scalings have been verified in our numerical simulations as can be seen in Fig.\ref{var-growth} and in Fig. 2 of the main text. Here $t_o$ is a constant to be determined from matching Ruelle's and Lorenz' regimes. From the condition that Kolmogorov-scale velocity perturbations are of order $\Delta(t_R) \sim (\varepsilon\nu)^{1/4}$ and the expressions for $\Delta (t)$ and $\ell(t),$ we deduce $t_o \sim t_R - (\nu/\varepsilon)^{1/2}.$ This implies that the randomization time $t_r(\ell)$ required for the stochastic wave to reach the scale $\ell$ is the sum of two terms: one proportional to the eddy-turnover time $(\ell/L)^{2/3}(L/U)\sim \varepsilon^{-1/3}\ell^{2/3}$ and the other dependent on the magnitude of thermal noise and proportional to $Re^{-1/2} \left(\mathrm{log} (\theta_\eta^{-1/2}) -1\right)(L/U).$ If the limit $Re^{-1},$ $\Theta \rightarrow 0$ is taken in any manner such that
	\be Re^{-1/2} \mathrm{log} \theta_\eta \rightarrow 0 \ee
	we arrive at the self-similar randomization time $t_r(\ell)\rightarrow C\varepsilon^{-1/3}\ell^{2/3}$, which is independent of the noise strength. The existence of a non-zero limit for $t_r(n)$ is the precise statement of spontaneous stochasticity. This result is consistent with earlier investigations of toy models of spontaneous stochasticity \cite{eyink2020renormalization, mailybaev2021spontaneously}, which suggest it is sufficient that the double limit be taken with $\Theta$ not decaying faster than $\mathrm{e}^{ - \gamma Re}$, where $\gamma$ is a constant. Such extreme robustness of spontaneous stochasticity is one of its characteristic features.}

\textcolor{black}{The above heuristic arguments based on ``eddy-turnover times'' at each length-scale are closely related to the spectrum of Lyapunov exponents 
	observed in chaotic shell models \cite{yamada1998asymptotic}, where local exponents $\lambda_n\sim \pm \varepsilon^{1/3} k_n^{2/3}$ are observed by 
	numerical calculations at each wavenumber $k_n$ in the inertial-range of scales. The existence of such a Lyapunov spectrum implies that the leading 
	(largest positive) Lyapunov exponent must diverge as $Re\to \infty.$ Such a spectrum of Lyapunov exponents thus constitutes a more refined criterion for existence of the ``inverse error cascade''  
	conjectured by Lorenz.}

\end{document}